\documentclass[a4paper,11pt]{article}
\usepackage{jheppub}
\usepackage{caption}
\usepackage{amsmath,amssymb}
\usepackage{subfig}
\usepackage{subfloat}
\usepackage{dsfont}
\usepackage{setspace}
\usepackage{hyperref}

\title{\boldmath Quasinormal Modes and GUP-Corrected Hawking Radiation of BTZ Black Holes within Modified Gravity Frameworks}

\author[a]{Faizuddin Ahmed}
\author[b]{,\,Ahmad Al-Badawi}
\author[c,1]{\,\.{I}zzet Sakall{\i}}
\author[d]{and Abdelmalek Bouzenada}

\affiliation[a]{Department of Physics, University of Science \& Technology Meghalaya, Ri-Bhoi, Meghalaya, 793101, India}
\affiliation[b]{Department of Physics, Al-Hussein Bin Talal University, 71111, Ma’an, Jordan}
\affiliation[c]{Physics Department, Eastern Mediterranean University, Famagusta 99628, North Cyprus via Mersin 10, Turkey}
\affiliation[d]{Laboratory of Theoretical and Applied Physics, Echahid Cheikh Larbi Tebessi University 12001, Algeria}

\emailAdd{faizuddinahmed15@gmail.com}
\emailAdd{ahmadbadawi@ahu.edu.jo}
\emailAdd{izzet.sakalli@emu.edu.tr}
\emailAdd{abdelmalekbouzenada@gmail.com}

\abstract{
This paper aims to explore the quasinormal modes (QNMs) and effective potential profiles of massless and rotating BTZ black holes within the frameworks of $f(\mathcal{R})$ and Ricci-Inverse ($\mathcal{RI}$) modified gravity theories, which, while producing similar space-time structures, exhibit variations due to distinct cosmological constants, $\Lambda_m$. We derive wave equations for these black hole perturbations and analyze the behavior of the effective potential $V_{\text{eff}}(r)$ under different values of mass $m$, cosmological constant $\Lambda_m$, and modified gravity parameters $\alpha_1$, $\alpha_2$, $\beta_1$, $\beta_2$, and $\gamma$. The findings indicate that increasing mass and parameter values results in a raised potential barrier, implying stronger confinement of perturbations and impacting black hole stability. Incorporating the generalized uncertainty principle, we also study its effect on the thermodynamics of rotating BTZ black holes, demonstrating how GUP modifies black hole radiation, potentially observable in QNM decay rates. Additionally, we investigate the motion of particles through null and timelike geodesics in static BTZ space-time, observing asymptotic behaviors for null geodesics and parameter-dependent shifts in potential for timelike paths. The study concludes that modified gravity parameters significantly influence QNM frequencies and effective potential profiles, offering insights into black hole stability and suggesting that these theoretical predictions may be tested through gravitational wave observations.}


\begin{document}
\maketitle
\flushbottom

\section{Introduction}\label{sec:1}

Einstein's general theory of relativity (GR) revolutionized our understanding of space-time and gravity, predicting the existence of black holes, regions of intense gravitational pull from which nothing, not even light, can escape. This prediction has gained substantial observational support, including the groundbreaking image of a black hole's shadow captured by the Event Horizon Telescope (EHT) collaboration in the galaxy M87 \cite{EventHorizonTelescope:2019dse}. This iconic observation aligns closely with $\text{GR}$ predictions and has sparked immense interest in exploring black hole physics, especially in studying black hole shadows and their behavior in various gravitational environments. Black holes, particularly supermassive ones, are thought to reside at galactic centers, making direct observation challenging. Yet, studying their interactions with surrounding matter and fields provides invaluable insights \cite{Ling:2021vgk, Yang:2022btw, Vagnozzi:2022moj, Jusufi:2019nrn, Haroon:2018ryd, Bambi:2019tjh, Vagnozzi:2019apd, Kumar:2020yem, Ghosh:2020ece, Adler:2022qtb}.

In galactic centers, black holes interact with accretion disks, dark matter, and magnetic fields, producing a variety of astrophysical phenomena, including the emission of gravitational waves. These gravitational waves, detected by observatories like LIGO, VIRGO, and the upcoming LISA mission, contain QNMs in their late stages, a signature characteristic of black holes. QNMs are essentially the "ringing" of space-time due to black hole perturbations, carrying information about the black hole's intrinsic properties, such as mass, charge, and angular momentum \cite{Konoplya:2011qq, Vishveshwara:1970zz}. The real part of QNM frequencies corresponds to oscillation rates, while the imaginary part indicates the decay rate, capturing black hole dynamics and stability \cite{Sun:2023woa, Zhang:2021bdr, Hendi:2020zyw, Jafarzade:2020ova, Anacleto:2021qoe, Lambiase:2023hng, Ghosh:2022gka, Pedrotti:2024znu, Ghosh:2023etd, Jusufi:2019ltj, Hamil:2024ppj}. Understanding QNMs is essential in exploring black hole stability and testing $\text{GR}$ under extreme conditions.

Despite its successes, $\text{GR}$ faces challenges at quantum scales and cosmological observations, prompting various modified gravity theories as potential extensions. $\mathcal{RI}$-gravity \cite{ref1,ref2,ref3} and $f(\mathcal{R})$-gravity \cite{bb3,bb4}  represent two such extensions that introduce corrections to $\text{GR}$ by incorporating higher-order terms of the Ricci scalar ($\mathcal{R}$), anti-curvature scalar ($\mathcal{A}$), and anti-curvature tensor ($A^{\mu \nu}$). In $\mathcal{RI}$-gravity, the function $f=f(\mathcal{R}, \mathcal{A}, A^{\mu \nu} A_{\mu \nu})$ extends the $\text{GR}$ Lagrangian to account for interactions influenced by these curvature terms, potentially affecting the space-time around compact objects like black holes. Recently, $\mathcal{RI}$-gravity has attracted much attention among researchers (see, for example, Refs. \cite{CJPHY,EPJC3,EPJC4,EPJP,NPB,AOP,JCAP} and related references there in). $f(\mathcal{R})$-gravity, on the other hand, consider modification to the Einstein-Hilbert action by replacing the Ricci scalar $\mathcal{R}$ by itself a function of it, $\mathcal{R} \to f(\mathcal{R})$. This altering the gravitational field equations. Such modified theories have found applications across cosmology, astrophysics, and theoretical investigations, aiming to address dark matter, dark energy, and quantum gravity issues.

In addition to modified gravity theories, quantum corrections to space-time structure are often modeled by the Generalized Uncertainty Principle (GUP) \cite{isMaggiore:1993rv,isScardigli:1999jh,isKonishi:1989wk,isHossenfelder:2012jw,isSakalli:2022swm,isSakalli:2023pgn,isKanzi:2019gtu}. The GUP introduces a minimal length scale, suggesting a modification to the Heisenberg uncertainty principle in high-energy regimes. In the context of black hole thermodynamics, GUP affects Hawking radiation, potentially slowing the rate of black hole evaporation and impacting the stability of small black holes \cite{CJPHY,isChen:2004ft,isGecim:2018sji,isPourhassan:2018chj,isLi:2018vgt}. 

Modified gravity theories represent a diverse class of theoretical frameworks designed to address the limitations of Einstein's GR, particularly in explaining phenomena such as dark energy, dark matter, and quantum gravitational effects. The present study focuses on the QNMs and the Hawking temperature in the context of modified gravity theories. Specifically, we investigate the QNMs of a stationary BTZ black hole within the frameworks of $\mathcal{RI}$-gravity and $f(\mathcal{R})$-gravity. Our objective is twofold: first, to analyze the QNMs associated with the BTZ black hole in these modified gravity settings, and second, to examine the GUP-corrected thermodynamics of the system. We show that the coupling constants related to the scalar curvature and its higher-order terms, the anti-curvature tensor, and its associated scalar contribute significantly to both the frequency of the QNMs and the GUP-corrected Hawking temperature. These modifications lead to results that differ from those predicted by GR, with notable shifts in both the QNM spectrum and the Hawking temperature. This study aims to provide a deeper understanding of how these modified gravity theories influence black hole dynamics and thermodynamics, and how the GUP corrections further alter the picture.

This paper is summarized as follows: In Sec. \ref{sec:2}, we discuss the BTZ space-time within these modified gravity theories, derive its properties, and analyze QNMs influenced by $\mathcal{RI}$-gravity and $f(\mathcal{R})$-gravity frameworks. Section \ref{sec:3} explores the GUP-modified thermodynamics of a rotating BTZ black hole, with a focus on Hawking radiation of bosons and fermions. In Sec. \ref{sec:4}, we present the geodesic structure for non-rotating BTZ black holes under modified gravity. Finally, Sec. \ref{sec:5} summarizes our conclusions and discusses potential implications of these results.

\section{BTZ space-time in modified gravity theories}\label{sec:2}

We begin this section by introducing the three-dimensional BTZ black hole rotating metric by the following line-element in the "Schwarzschild" coordinates ($t, r, \psi$) \cite{r5,r6}
\begin{eqnarray}
ds^2=g_{tt}\,dt^2+2\,g_{t\psi}\,dt\,d\psi+g_{rr}\,dr^2+g_{\psi\psi}\,d\psi^2,\label{a1}
\end{eqnarray}
where different coordinates are in the ranges $-\infty < t=x^0 < +\infty$, $0 < r=x^1 < +\infty$, and $\psi=x^3 \sim \psi+2\,n\,\pi$ ($n=\pm\,1,\pm\,2,....$). The non-zero components of the metric tensor $g_{\mu\nu}$ are $(\mu,\nu=0,1,2)$
\begin{eqnarray}
g_{\mu\nu}&=&\left(\begin{array}{ccc}
    M+\Lambda\,r^2 & 0 & -\frac{J}{2}\\
     0 & \frac{1}{\Big(-M-\Lambda\,r^2+\frac{J^2}{4\,r^2}\Big)} & 0\\
     -\frac{J}{2} & 0 & r^2
\end{array}\right).\label{a2}
\end{eqnarray}
The two constants of integration, $M$ and $J$, correspond to the conserved charges associated with asymptotic invariance under time displacements (representing mass) and rotational invariance (representing angular momentum), respectively.

Now, we discuss $(2+1)$-dimensional black hole solutions in the modified gravity theories and, in the sequel, we shall investigate the physical features of the obtained space-time. The first one being $\mathcal{RI}$-gravity and next one is the $f(\mathcal{R})$-gravity.

The action that describes the $\mathcal{RI}$-gravity is given by \cite{ref1,ref2,ref3,arxiv}
\begin{eqnarray}
    \mathcal{S}= \int \sqrt{-g}\left[f(\mathcal{R}, \mathcal{A}, A^{\mu\nu}\,A_{\mu\nu})-2\,\Lambda\right]\,dt\,d^2x++{\cal S}_m,\label{a3}
\end{eqnarray}
where ${\cal S}_m$ is the action of matter Lagrangian, and other symbols have their usual meanings. 

By varying the action (\ref{a3}) with respect to the metric tensor $g_{\mu\nu}$, the modified field equations is given by  
\begin{equation}\label{a4}
    -\frac{1}{2}\, f\, g^{\mu \nu} + f_{\mathcal{R}} \, R^{\mu \nu}-f_{\mathcal{A}} \, A^{\mu \nu} - 2\,f_{A^2}\, A^{\rho \nu }A^{\mu}_{\rho} +P^{\mu \nu}+ M^{\mu \nu} + U^{\mu \nu} + \Lambda_m g^{\mu \nu} =\mathcal{T}^{\mu \nu},
\end{equation}
where $\Lambda_m$ is the cosmological constant in this theory and other tensors are as follows:
\begin{eqnarray}
    P^{\mu \nu} &=& g^{\mu \nu} \nabla ^2 \, f_{\mathcal{R}}-\nabla ^{\mu} \nabla^{\nu}\, f_{\mathcal{R}} \, ,\label{a5}\\ 
    M^{\mu \nu} &=& g^{\rho \mu}\nabla _{\alpha } \nabla _{\rho } (f_{\mathcal{A}}\, A_{\sigma}^{ \alpha}\, A^{\nu \sigma}) - \frac{1}{2}\, \nabla ^2 (f_{\mathcal{A}}\,A^{\mu}_{\sigma}\, A^{\nu \sigma}) - \frac{1}{2}\, g^{\mu \nu}\, \nabla _{\alpha} \nabla _{ \rho} ( f_{\mathcal{A}}\,A_{\sigma}^{ \alpha}\, A^{\rho \sigma})\, ,\label{a6}\\
 \nonumber  U^{\mu \nu} &=& g^{\rho \nu}\,\nabla _{\alpha} \nabla _{\rho}(f_{A^2}\,A_{\sigma \kappa}\,A^{\sigma \alpha}\,A^{\mu \kappa})-\nabla ^{2}(f_{A^2}\,A_{\sigma \kappa}\,A^{\sigma \mu}\,A^{\nu \kappa})
  -g^{\mu \nu}\,\nabla _{\alpha} \nabla _{\rho}(f_{A^2}\,A_{\sigma \kappa}\,A^{\sigma \alpha}\,A^{\rho \kappa})\nonumber\\
  &+& 2\,g^{\rho \nu}\,\nabla _{\rho} \nabla _{\alpha}(f_{A^2}\,A_{\sigma \kappa}\,A^{\sigma \mu}\,A^{\alpha \kappa})
   - g^{\rho \nu}\,\nabla _{\alpha} \nabla _{\rho}(f_{A^2}\,A_{\sigma \kappa}\,A^{\sigma \mu}\,A^{\alpha \kappa})\, .\label{a7} 
\end{eqnarray}
Here various symbols are defined by
\begin{equation}
    f_{\mathcal{R}}=\partial f/\partial \mathcal{R},\quad f_{\mathcal{A}}=\partial f/\partial \mathcal{A},\quad f_{A^2}=\partial f/\partial (A^{\mu\nu}\,A_{\mu\nu})\,.\label{a8}
\end{equation}

Let's consider the function $f$ in the general Class to be the following form:
\begin{equation} \label{a9}
    f(\mathcal{R},\mathcal{A},A^{\mu \nu}\,A_{\mu \nu}) =\mathcal{R}+{\alpha_1\,\mathcal{R}^2}+\alpha_2\,\mathcal{R}^3+\beta_1\, \mathcal{A}+\beta_2\,\mathcal{A}^2+\gamma\,A^{\mu \nu}\,A_{\mu \nu} \, 
\end{equation}
with $\alpha_i, \beta_i$\quad ($i=1,2$) and $\gamma$ being arbitrary constants.

Solving the modified field equations (\ref{a4}) with the energy-momentum tensor $\mathcal{T}^{\mu \nu}=0$ and after simplification, we obtain the effective cosmological constant given by \cite{arxiv}
\begin{equation}
    \Lambda^{\mathcal{RI}}_m=\Lambda-6\,\alpha_1\,\Lambda^2-108\,\alpha_2\,\Lambda^3+\frac{5\,\beta_1}{4\,\Lambda}+\frac{15\,\beta_2}{8\,\Lambda^2}+\frac{7\,\gamma}{8\,\Lambda^2}.\label{a10}
\end{equation}
We see that the cosmological constant gets modification by the coupling constants $(\alpha_i,\beta_i,\gamma)$ compared to GR result. The BTZ metric in $\mathcal{RI}$-gravity is therefore given by Eq. \eqref{a1} with the following metric tensor:
\begin{eqnarray}
g_{\mu\nu}=\left(\begin{array}{ccc}
     M+\Lambda_m\,r^2 & 0 & -\frac{J}{2}  \\
     0 & \frac{1}{\Big(-M-\Lambda_m\,r^2+\frac{J^2}{4\,r^2}\Big)} & 0\\
     -\frac{J}{2} & 0 & r^2
\end{array}\right),\label{special-metric2}
\end{eqnarray}
with $\Lambda_m \to \Lambda^{\mathcal{RI}}_m$ in the context of $RI$-gravity as given in Eq. (\ref{a10}).

Next we study another modified theory known as $f(\mathcal{R})$-gravity. Therefore, the Lagrangian that describes $f(\mathcal{R})$-gravity theory is given by
\begin{eqnarray}
    S= \int dt\,d^2x\, \sqrt{-g}\left[f(\mathcal{R})-2\,\Lambda_m+{\cal L}_m\right],\label{ss1}
\end{eqnarray}

Varying the action with respect to the metric tensor results the modified field equations for $f(\mathcal{R})$-gravity with the cosmological constant given by 
\begin{equation}\label{ss2}
    f_{\mathcal{R}}\,R^{\mu \nu}-\frac{1}{2}\, f(\mathcal{R})\, g^{\mu \nu}-\nabla^{\mu}\, \nabla^{\nu} \, f_{\mathcal{R}} + g^{\mu \nu}\, \nabla^{\alpha}\, \nabla_{\alpha} \, f_{\mathcal{R}} + \Lambda _m\, g^{\mu \nu}=\mathcal{T}^{\mu \nu}\, 
\end{equation}

In this analysis, we choose the function $f(\mathcal{R})$ to be the following form
\begin{equation} \label{ss3}
    f(\mathcal{R})=\mathcal{R}+{\alpha_1\,\mathcal{R}^2}+\alpha_2\,\mathcal{R}^3 
\end{equation}
where $\alpha_1, \alpha_2$ are the coupling constant.

By solving the field equations (\ref{ss2}) for zero energy-momentum tensor, $\mathcal{T}^{\mu\nu}=0$ and after simplification, we obtain the effective cosmological constant given as follows \cite{arxiv}:
\begin{equation}
    \Lambda^{f(\mathcal{R})}_m=\Lambda-6\,\alpha_1\,\Lambda^2-108\,\alpha_2\,\Lambda^3.\label{ss4}
\end{equation}
that reduces to GR result for $\alpha_1=0=\alpha_2$. Consequently, in $f(\mathcal{R})$-gravity, the BTZ metric is described by Eq. \eqref{a1}, incorporating the metric tensor specified below:
\begin{eqnarray}
g_{\mu\nu}=\left(\begin{array}{ccc}
     M+\Lambda_m\,r^2 & 0 & -\frac{J}{2}  \\
     0 & \frac{1}{\Big(-M-\Lambda_m\,r^2+\frac{J^2}{4\,r^2}\Big)} & 0\\
     -\frac{J}{2} & 0 & r^2
\end{array}\right),\label{ss6}
\end{eqnarray}
with $\Lambda_m \to \Lambda^{f(\mathcal{R})}_m$ in the context of $f(\mathcal{R})$ as given in Eq. (\ref{ss4}). In the following sections, we study QNMs for the metrics (\ref{special-metric2}) and (\ref{ss6}), and analyze the results.

\subsection{QNMs of massless BTZ space-time in modified gravity}\label{sec:2.1}

We discuss the scalar field perturbation of massless BTZ space-time within the frameworks of $f(\mathcal{R})$ and $\mathcal{RI}$-gravity and search for the corresponding effective potential as well as QNM frequency. Thus, massless and non-rotating BTZ space-time ($M=0=J$) is described by the following line-element
\begin{eqnarray}
ds^2=g_{tt}\,dt^2+g_{rr}\,dr^2+g_{\psi\psi}\,d\psi^2,\label{special-metric3}
\end{eqnarray}
where the metric tensor $g_{\mu\nu}$ is given by 
\begin{eqnarray}
g_{\mu\nu}=\left(\begin{array}{ccc}
     -(-\Lambda_m)\,r^2 & 0 & 0  \\
     0 & \frac{1}{(-\Lambda_m)\,r^2} & 0\\
     0 & 0 & r^2
\end{array}\right).\label{special-metric4}
\end{eqnarray}

For a massless scalar field $\Phi$, the equation of motion is the general covariant Klein-Gordon equation given by \cite{LL} 
\begin{equation}
    \frac{1}{\sqrt{-g}}\,\partial_{\mu}\,\left(\sqrt{-g}\,g^{\mu\nu}\,\partial_{\nu}\Phi\right)=0.\label{C1}
\end{equation}

Given that the background space-time (\ref{special-metric3}) is axially symmetric and static, we can express the massless scalar field function $\Phi(t,r,\psi)$ as a decomposition in terms of $\mathrm{R}(r)$ as follows:
\begin{equation}
    \Phi(t,r,\psi)=\exp(-i\,\omega\,t)\,\exp(i\,m\,\psi)\,\frac{\mathrm{R}(r)}{\sqrt{r}},\label{C2}
\end{equation}
where $\omega$ is the frequency of oscillation and $m$ takes the natural number.

Expressing the wave equation (\ref{C1}) in the background (\ref{special-metric3}) and substituting relation (\ref{C2}), we obtain the following differential equation 
\begin{equation}
    r^4\,\Lambda^2_{m}\,\frac{d^2\mathrm{R}}{dr^2}+2\,r^3\,\Lambda^2_{m}\,\frac{d\mathrm{R}}{dr}+\left(\omega^2+m^2\,\Lambda_m-\frac{3}{4}\,r^2\,\Lambda^2_{m}\right)\,\mathrm{R}=0.\label{C3}
\end{equation}
Now, defining a new coordinate (called tortoise coordinate) as follows:
\begin{equation}
    r_*=\int\,\frac{dr}{(-\Lambda_m)\,r^2}=-\frac{1}{(-\Lambda_m)\,r}\label{C4}
\end{equation}

Using this new coordinate into the differential equation (\ref{C3}), we obtain a Schr\"{o}dinger-like wave equation form given by
\begin{equation}
    \frac{d^2\,\mathrm{R}}{dr^{2}_{*}}+\left(\omega^2-V_{eff}\right)\,\mathrm{R}=0.\label{C5}
\end{equation}
where the effective potential $V_{eff}$ in the wave equation (\ref{C5}) is given as follows
\begin{equation}
    V_{eff}(r)=-m^2\,\Lambda_m+\frac{3}{4}\,r^2\,\Lambda^2_{m}.\label{C6}
\end{equation}

The effect of variations of coupling constants $\alpha_i, \beta_i$ and $\gamma$ on the effective potential $V_{eff}$ is illustrated in Figure \ref{fig:1}. The significant result on the effective potential is observed when the value of these coupling constants gradually increases for a particular $m$-state and compared with the GR result. In panel (a) of Figure \ref{fig:1}, the dotted red line representing $m=0$ corresponds to the GR case, where the parameters are set to $\alpha_i=0$, $\beta_i=0$, and $\gamma=0$. In contrast, the purple line for $m=0$ represents the $\mathcal{RI}$-gravity scenario, with the coupling constants set as $\alpha_1=0.5$, $\alpha_2=0.001$, $\beta_1=0.1$, $\beta_2=0.001$, and $\gamma=0.01$. The other lines in this panel depict different values of  $m$ while maintaining the same coupling constants. In panel (b) of Figure \ref{fig:1}, the lines correspond to various values of the coupling constants with $m=0$ state. This setup allows for a comparative analysis of the effective potential for a massless scalar field in massless BTZ space-time within the framework of GR and $\mathcal{RI}$-gravity theories.

Similarly, we have generated Figure \ref{fig:2} showing the effective potential of QNMs system within the framework of $f(\mathcal{R})$ gravity on massless BTZ space-time. The significant result on the effective potential is observed when the value of these coupling constants gradually increases for a particular $m$-state and compared with the GR result. In panel (a) of Figure \ref{fig:2}, the dotted red line representing $m=0$ corresponds to the GR case, where the parameters are set to $\alpha_i=0$. In contrast, the purple line for $m=0$ represents $f(\mathcal{R})$-gravity scenario, with the coupling constants set as $\alpha_1=0.5$, and $\alpha_2=0.001$. The other lines in this panel depict different values of  $m$ while maintaining the same coupling constants. In panel (b) of Figure \ref{fig:2}, the lines correspond to various values of the coupling constants with $m=0$ state. This setup allows for a comparative analysis of the effective potential for massless scalar field in massless BTZ space-time within the framework of GR and $f(\mathcal{R})$-gravity theories.

\begin{center}
\begin{figure}[ht!]
\begin{centering}
\subfloat[$\Lambda=-0.1$]{\centering{}\includegraphics[scale=0.65]{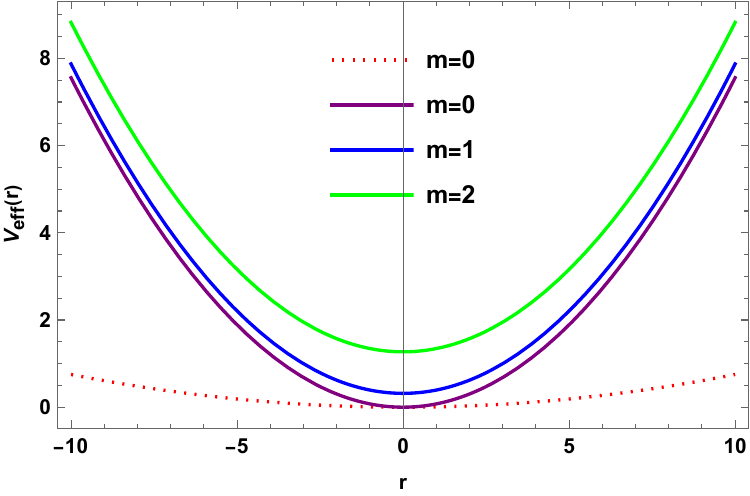}}\quad\quad
\subfloat[$\Lambda=-0.1$, $m=0$]{\centering{}\includegraphics[scale=0.65]{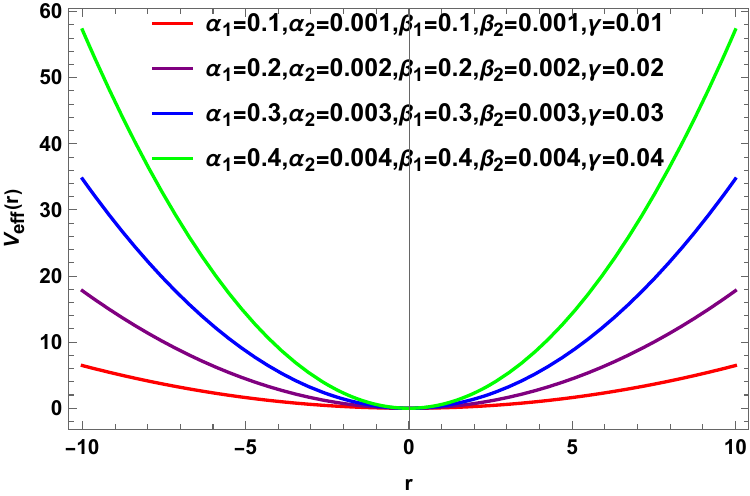}}\quad
\centering{}\caption{The effective potential (\ref{C6}) on massless BTZ space-time in RI gravity.}\label{fig:1}
\hfill\\
\subfloat[$\Lambda=-0.1$]{\centering{}\includegraphics[scale=0.65]{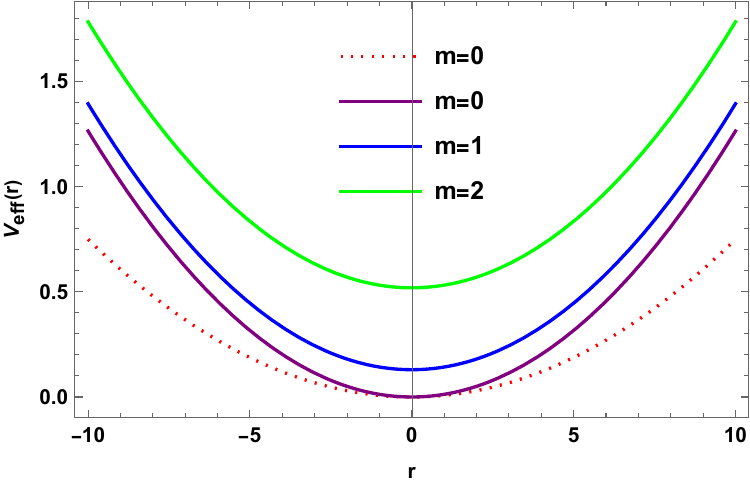}}\quad\quad
\subfloat[$\Lambda=-0.1$, $m=0$]{\centering{}\includegraphics[scale=0.65]{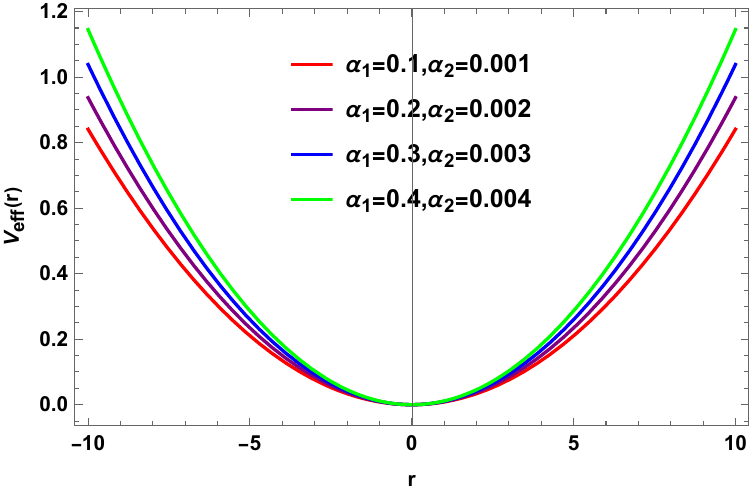}}\quad
\centering{}\caption{The effective potential (\ref{C6}) on massless BTZ space-time in $f(\mathcal{R})$ gravity.}\label{fig:2}
\end{centering}
\end{figure}
\par\end{center}

Based on the effective potential obtained above, we can study the QNM frequencies of massless BTZ space-time. One can calculate the QNMs using the WKB method and investigate its associated properties. It is important to note that QNM emerges when appropriate boundary conditions are applied before performing the calculations
\begin{equation}
    \Phi(r_{*}) \sim e^{\pm\,i\,\omega\,r_*},\quad\quad r_{*} \to \pm\,\infty,\label{C7}
\end{equation}
where purely ingoing modes occur at $r_{*}=-\infty$ and purely outgoing
modes occur at $r_{*}=\infty$.

In terms of $r_{*}$, one can write the effective potential as follows:
\begin{equation}
    V_{eff}(r_{*})=-m^2\,\Lambda_m+\frac{s\,(s+1)}{r^2_{*}},\quad\quad s=1/2.\label{C8}
\end{equation}
Since at $r \to 0$, $r_{*} \to -\infty$ and $r\to \infty$, one will $r_{*} \to 0$, there is only purely ingoing modes occur. 

Equation (\ref{C5}) therefore can be expressed as follows:
\begin{equation}
    \frac{d^2\,\mathrm{R}}{dr^{2}_{*}}+\left(\tilde{\omega}^2-\frac{s\,(s+1)}{r^2_{*}}\right)\,\mathrm{R}=0,\label{C9}
\end{equation}
where
\begin{equation}
    \tilde{\omega}=\sqrt{\omega^2+m^2\,\Lambda_m}>0.\label{C10}
\end{equation}
Equation (\ref{C9}) can be solved through special functions, like Bessel functions.

Transforming Equation (\ref{C9}) by a new variable $\rho=\tilde{\omega}\,r_{*}$, we obtain
\begin{equation}
    \frac{d^2\,\mathrm{R}}{d\rho^2}+\left[1-\frac{s\,(s+1)}{\rho^2}\right]\,\mathrm{R}=0.\label{C11}
\end{equation}
The solution of Eq. (\ref{C11}) is given by
\begin{equation}
    \mathrm{R}(\rho)=\rho\,(c_1\,j_{s} (\rho)-c_2\,Y_{s} (\rho)),\label{C12}
\end{equation}
where $c_i$ are arbitrary constants and $j_{s}(\rho), Y_{s}(\rho)$ are spherical Bessel and spherical Neumann functions respectively. Note that at $\rho=0$, $\mathrm{R}(\rho=0)=0$ which implies $c_2=0$. Thus, the final solution is given by
\begin{equation}
    \mathrm{R}(r_{*})=c_1\,\tilde{\omega}\,r_{*}\,j_{s} (\tilde{\omega}\,r_{*}).\label{C13}
\end{equation}

Therefore, the massless scalar field solution is given by
\begin{equation}
    \Phi(t,r_{*},\psi)=c_1\,\sqrt{\omega^2+m^2\,\Lambda_m}\,\exp(-i\,\omega\,t)\,\exp(i\,m\,\psi)\,\sqrt{r_{*}}\,j_{s} (\sqrt{\omega^2+m^2\,\Lambda_m}\,r_{*}),\label{C14}
\end{equation}
where $\omega >|m|\,\sqrt{-\Lambda_m}$.

For $\rho>>s$, we can write asymptotic form of the spherical Bessel function
\begin{equation}
    j_{s} (\rho) \sim \frac{1}{\rho}\,\sin \left(\rho-\frac{s\,\pi}{2}\right).\label{C15}
\end{equation}
Thus, the equation (\ref{C13}) in asymptotic form can be expressed as
\begin{equation}
    \mathrm{R}(\rho) \sim \sin \left(\rho-\frac{s\,\pi}{2}\right).\label{C16}
\end{equation}
In terms of $r_{*}$, we can write Eq. (\ref{C16}) as follows:
\begin{equation}
    \mathrm{R}(r_{*}) \sim \sin \left(\tilde{\omega}\,r_{*}-\frac{s\,\pi}{2}\right).\label{C17}
\end{equation}

At $r_{*}=\bar{r}_{*}>0$, that there is a certain values of the tortoise coordinate $r_{*}$ for which this radial function vanishes, $\mathrm{R}(\bar{r}_{*})=0$. This implies
\begin{eqnarray}
    &&\sin \left(\tilde{\omega}\,\bar{r}_{*}-\frac{s\,\pi}{2}\right)=0\Rightarrow
    \tilde{\omega}\,\bar{r}_{*}=\left(2\,n+1+\frac{s}{2}\right)\,\pi\nonumber\\\Rightarrow
   &&\omega_{n\,m}=\sqrt{m^2\,(-\Lambda_m)+\left(2\,n+\frac{5}{4}\right)^2\,\left(\frac{\pi}{\bar{r}_{*}}\right)^2},\label{C18}
\end{eqnarray}
where $\Lambda_m \to \Lambda^{\mathcal{RI}}_m$ in $RI$-gravity as given in Eq. (\ref{a10}) and $\Lambda_m \to \Lambda^{f(\mathcal{R})}_m$ in $f(\mathcal{R})$-gravity as given in Eq. (\ref{ss4}).

From the above analysis, we see that the frequency of QNMs $\omega_{n\,m}$ with $m \neq 0$ depends on the coupling constants $\alpha_i, \beta_i$ and $\gamma$ in $\mathcal{RI}$ gravity and $\alpha_i$ in the context of $f(\mathcal{R})$-gravity.

\subsection{QNMs of $AdS_3$-type BTZ space-time in modified gravity}\label{sec:2.2}

We discuss the scalar field perturbation of $AdS_3$-type BTZ space-time within the frameworks of $f(\mathcal{R})$ and $\mathcal{RI}$-gravity and search for the corresponding effective potential as well as QNM frequency. Thus, $AdS_3$-type BTZ space-time ($M=-1,J=0$) is described by the following line-element
\begin{eqnarray}
ds^2=g_{tt}\,dt^2+g_{rr}\,dr^2+g_{\psi\psi}\,d\psi^2,\label{special-metric5}
\end{eqnarray}
where the metric tensor $g_{\mu\nu}$ is given by 
\begin{eqnarray}
g_{\mu\nu}=\left(\begin{array}{ccc}
     -(1-\Lambda_m\,r^2) & 0 & 0  \\
     0 & \frac{1}{(1-\Lambda_m\,r^2)} & 0\\
     0 & 0 & r^2
\end{array}\right),\quad g^{\mu\nu}=\left(\begin{array}{ccc}
     -(1-\Lambda_m\,r^2)^{-1} & 0 & 0  \\
     0 & (1-\Lambda_m\,r^2) & 0\\
     0 & 0 & r^{-2}
\end{array}\right).\label{special-metric6}
\end{eqnarray}

Expressing the wave equation (\ref{C1}) in the background (\ref{special-metric5}) and substituting relation (\ref{C2}), we obtain the following differential equation 
\begin{equation}
    (1-r^2\,\Lambda_{m})\,\frac{d^2\mathrm{R}}{dr^2}+2\,r\,(-\Lambda_{m})\,\frac{d\mathrm{R}}{dr}+\left[\frac{\omega^2}{(1-r^2\,\Lambda_m)}-\frac{\left(m^2-\frac{1}{4}\,(1+3\,r^2\,\Lambda_m)\right)}{r^2}\right]\,\mathrm{R}=0.\label{D1}
\end{equation}
Now, defining a new coordinate (called tortoise coordinate) as follows:
\begin{equation}
    r_*=\int\,\frac{dr}{(1-\Lambda_m\,r^2)}=\frac{1}{\sqrt{-\Lambda_m}}\,\tan^{-1}(\sqrt{-\Lambda_m}\,r)\Rightarrow \sqrt{-\Lambda_m}\,r=\tan(\sqrt{-\Lambda_m}\,r_*).\label{D2}
\end{equation}

With this coordinate transformation, we can rewrite the above differential equation (\ref{D1}) as follows:
\begin{equation}
    \frac{d^2\,\mathrm{R}}{dr^{2}_{*}}+\left(\omega^2-V_{eff}\right)\,\mathrm{R}=0.\label{D3}
\end{equation}
where the effective potential $V_{eff}$ in the wave equation (\ref{D3}) is given as follows
\begin{equation}
    V_{eff}(r)=\frac{(1-r^2\,\Lambda_m)}{r^2}\,\left[m^2-\frac{1}{4}\,(1+3\,r^2\,\Lambda_m)\right].\label{D4}
\end{equation}

\begin{center}
\begin{figure}[ht!]
\begin{centering}
\subfloat[$\Lambda=-0.1$]{\centering{}\includegraphics[scale=0.65]{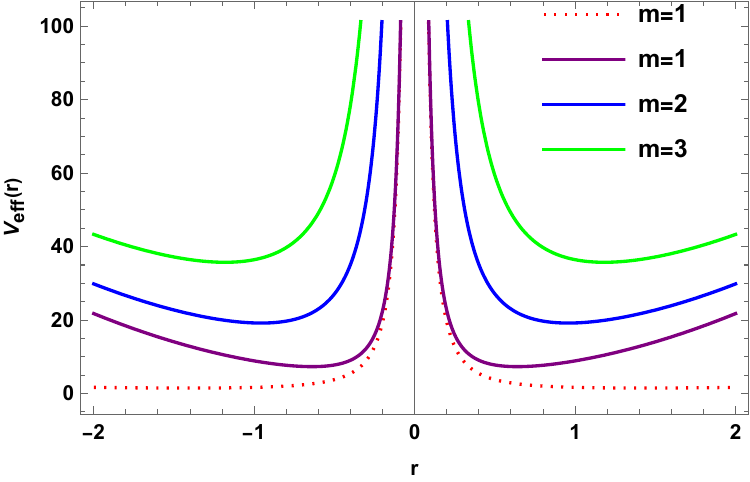}}\quad\quad
\subfloat[$\Lambda=-0.1$, $m=0$]{\centering{}\includegraphics[scale=0.65]{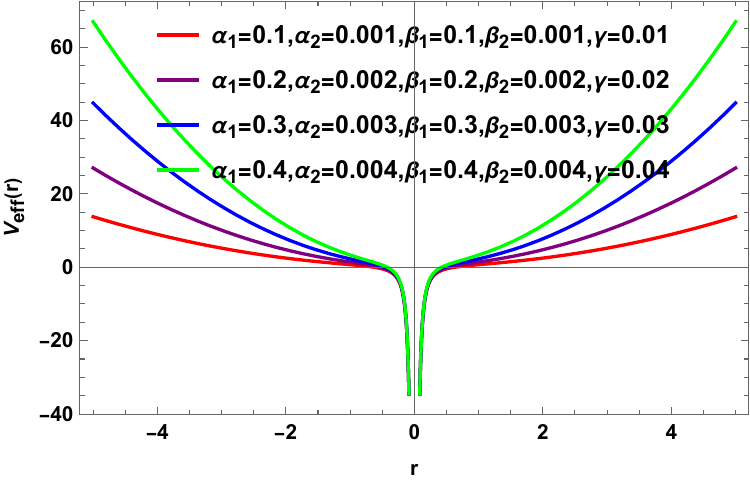}}\quad
\centering{}\caption{The effective potential (\ref{D4}) on AdS-type BTZ space-time in RI gravity.}\label{fig:3}
\hfill\\
\subfloat[$\Lambda=-0.1$]{\centering{}\includegraphics[scale=0.65]{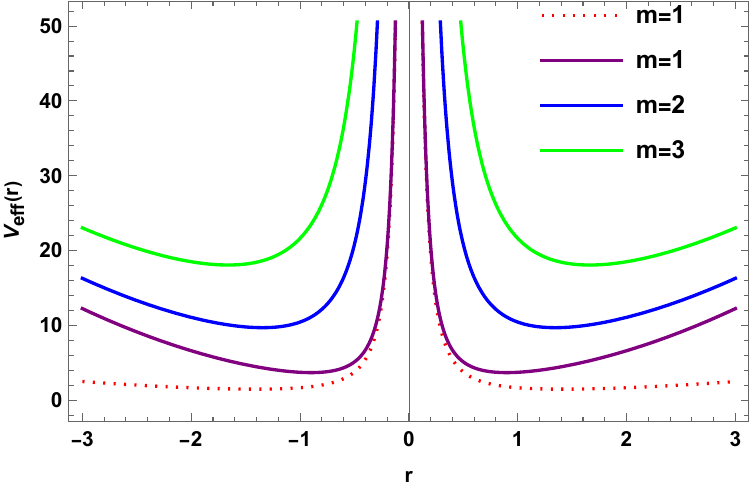}}\quad\quad
\subfloat[$\Lambda=-0.1$, $m=0$]{\centering{}\includegraphics[scale=0.65]{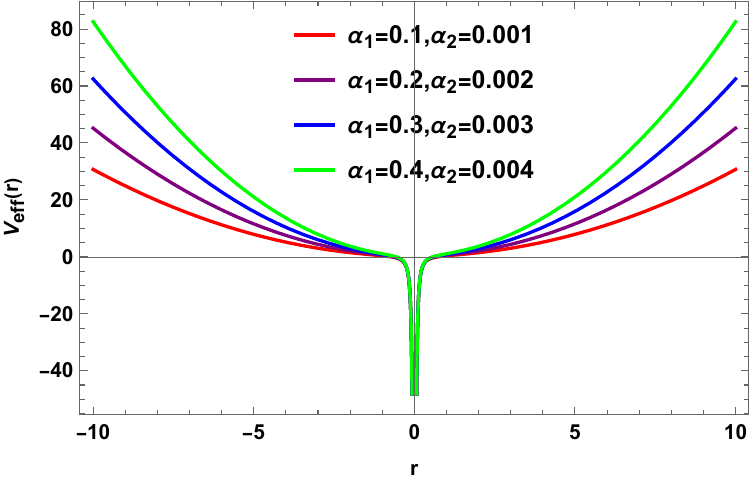}}\quad
\centering{}\caption{The effective potential (\ref{D4}) on AdS-type BTZ space-time in $f(\mathcal{R})$ gravity.}\label{fig:4}
\end{centering}
\end{figure}
\par\end{center}

The effect of variations of coupling constants $\alpha_i, \beta_i$ and $\gamma$ on the effective potential $V_{eff}$ is illustrated in Figure \ref{fig:3}. The significant result on the effective potential is observed when the value of these coupling constants gradually increases for a particular $m$-state and compared with the GR result. In panel (a) of Figure \ref{fig:3}, the dotted red line representing $m=1$ corresponds to the GR case, where the parameters are set to $\alpha_i=0$, $\beta_i=0$, and $\gamma=0$. In contrast, the purple line for $m=1$ represents the $\mathcal{RI}$-gravity scenario, with the coupling constants set as $\alpha_1=0.5$, $\alpha_2=0.001$, $\beta_1=0.5$, $\beta_2=0.001$, and $\gamma=0.01$. The other lines in this panel depict different values of  $m$ while maintaining the same coupling constants. In panel (b) of Figure \ref{fig:3}, the lines correspond to various values of the coupling constants with $m=0$ state.

Similarly, we have generated Figure \ref{fig:4} showing the effective potential of QNMs system within the framework of $f(\mathcal{R})$ gravity on AdS-type BTZ space-time. The significant result on the effective potential is observed when the value of these coupling constants gradually increases for a particular $m$-state and compared with the GR result. In panel (a) of Figure \ref{fig:2}, the dotted red line representing $m=0$ corresponds to the GR case, where the parameters are set to $\alpha_i=0$. In contrast, the purple line for $m=1$ represents $f(\mathcal{R})$-gravity scenario, with the coupling constants set as $\alpha_1=0.5$, and $\alpha_2=0.001$. The other lines in this panel depict different values of  $m$ while maintaining the same coupling constants. In panel (b) of Figure \ref{fig:4}, the lines correspond to various values of the coupling constants with $m=0$ state.

In terms of $r_*$, we obtain the following expression of the effective potential given by
\begin{equation}
    V_{eff}(r_*)=(-\Lambda_m)\,\left[\frac{(m^2-1/4)}{\sin^2(\sqrt{-\Lambda_m}\,r_*)}+\frac{3/4}{\cos^2(\sqrt{-\Lambda_m}\,r)}\right].\label{D5}
\end{equation}

Equation (\ref{D3}) therefore can be expressed as follows:
\begin{equation}
    \frac{d^2\,\mathrm{R}}{dr^{2}_{*}}+\left[\omega^2-\frac{(m^2-1/4)\,\kappa^2}{\sin^2(\kappa\,r_*)}-\frac{(3/4)\,\kappa^2}{\cos^2(\kappa\,r_*)}\right]\,\mathrm{R}=0,\quad\quad \kappa=\sqrt{-\Lambda_m}.\label{D6}
\end{equation}
Transforming $y=\kappa\,r_*$, we obtain
\begin{equation}
    \frac{d^2\,\mathrm{R}}{dy}+\left(\mathrm{a}-\frac{\mathrm{b}}{\sin^2 y}-\frac{\mathrm{c}}{\cos^2 y}\right)\,\mathrm{R}=0,\label{D7}
\end{equation}
where
\begin{equation}
    \mathrm{a}=\frac{\omega^2}{\kappa^2},\quad\quad\mathrm{b}=m^2-1/4,\quad\quad \mathrm{c}=3/4.\label{D8}
\end{equation}

The above differential equation (\ref{D7}) can easily be solved using special functions and we find
\begin{equation}
    \omega_{n\,m}=2\,\sqrt{-\Lambda_m}\,\left(n+1+\frac{|m|}{2}\right),\label{D9}
\end{equation}
where $\Lambda_m \to \Lambda^{\mathcal{RI}}_m$ in $RI$-gravity as given in Eq. (\ref{a10}) and $\Lambda_m \to \Lambda^{f(\mathcal{R})}_m$ in $f(\mathcal{R})$-gravity as given in Eq. (\ref{ss4}).

From the above analysis, we see that the frequency of QNMs $\omega_{n\,m}$ depends on the coupling constants $\alpha_i, \beta_i$ and $\gamma$ in $\mathcal{RI}$ gravity and $\alpha_i$ in the context of $f(\mathcal{R})$-gravity.

\subsection{QNMs of massive non-rotating BTZ space-time in modified gravity}\label{sec:2.3}

Here, we discuss the scalar field perturbation of massive non-rotating BTZ space-time within the frameworks of $f(\mathcal{R})$ and $\mathcal{RI}$-gravity and search for the corresponding effective potential as well as frequency of QNMs. Thus, non-rotating BTZ space-time ($J=0$) is described by the following line-element
\begin{eqnarray}
ds^2=g_{tt}\,dt^2+g_{rr}\,dr^2+g_{\psi\psi}\,d\psi^2,\label{special-metric7}
\end{eqnarray}
where the metric tensor $g_{\mu\nu}$ is given by 
\begin{eqnarray}
g_{\mu\nu}=\left(\begin{array}{ccc}
     (M+\Lambda_m\,r^2) & 0 & 0  \\
     0 & \frac{1}{-(M+\Lambda_m\,r^2)} & 0\\
     0 & 0 & r^2
\end{array}\right),\quad g^{\mu\nu}=\left(\begin{array}{ccc}
     (M+\Lambda_m\,r^2)^{-1} & 0 & 0  \\
     0 & -(M+\Lambda_m\,r^2) & 0\\
     0 & 0 & r^{-2}
\end{array}\right).\label{special-metric8}
\end{eqnarray}

Expressing the wave equation (\ref{C1}) in the background (\ref{special-metric7}) and substituting relation (\ref{C2}), we obtain the following differential equation 
\begin{equation}
    (-M-r^2\,\Lambda_{m})\,\frac{d^2\mathrm{R}}{dr^2}+2\,(-\Lambda_{m})\,r\,\frac{d\mathrm{R}}{dr}+\left[\frac{\omega^2}{(-M-r^2\,\Lambda_{m})}-\frac{\left(m^2-\frac{1}{4}\,(-M+3\,r^2\,\Lambda_m)\right)}{r^2}\right]\,\mathrm{R}=0.\label{EE1}
\end{equation}
Now, defining a new coordinate (called tortoise coordinate) as follows:
\begin{equation}
    r_*=\int\,\frac{dr}{-M-\Lambda_m\,r^2}=\left\{\,\begin{array}{c}
        \frac{1}{\,\sqrt{M\,\Lambda_m}}\,\tan^{-1}\left(\sqrt{\frac{\Lambda_m}{M}}\,r\right),\quad |M|<0  \\
         \frac{1}{2\,\sqrt{M\,(-\Lambda_m)}}\,\mbox{ln}\left|\frac{r-\sqrt{\frac{M}{-\Lambda_m}}}{r+\sqrt{\frac{M}{-\Lambda_m}}}\right|,\quad\quad |M|>0 
    \end{array}
    \right.
    \label{EE2}
\end{equation}

With this coordinate transformation, we can rewrite the above differential equation (\ref{D1}) as follows:
\begin{equation}
    \frac{d^2\,\mathrm{R}}{dr^{2}_{*}}+\left(\omega^2-V_{eff}\right)\,\mathrm{R}=0.\label{EE3}
\end{equation}
where the effective potential $V_{eff}$ in the wave equation (\ref{D3}) is given as follows
\begin{equation}
    V_{eff}(r)=\frac{1}{r^2}\,(-M-r^2\,\Lambda_m)\,\left(m^2-\frac{1}{4}\,(-M+3\,r^2\,\Lambda_m)\right).\label{EE4}
\end{equation}
where $\Lambda_m \to \Lambda^{\mathcal{RI}}_m$ in $RI$-gravity as given in Eq. (\ref{a10}) and $\Lambda_m \to \Lambda^{f(\mathcal{R})}_m$ in $f(\mathcal{R})$-gravity as given in Eq. (\ref{ss4}).

From the above analysis, we see that the frequency of QNMs $\omega_{n\,m}$ depends on the coupling constants $\alpha_i, \beta_i$ and $\gamma$ in $\mathcal{RI}$ gravity and $alpha_i$ in the context of $f(\mathcal{R})$-gravity.

\begin{center}
\begin{figure}[htb!]
\subfloat[$\Lambda=-0.1,M=2$]{\centering{}\includegraphics[scale=0.65]{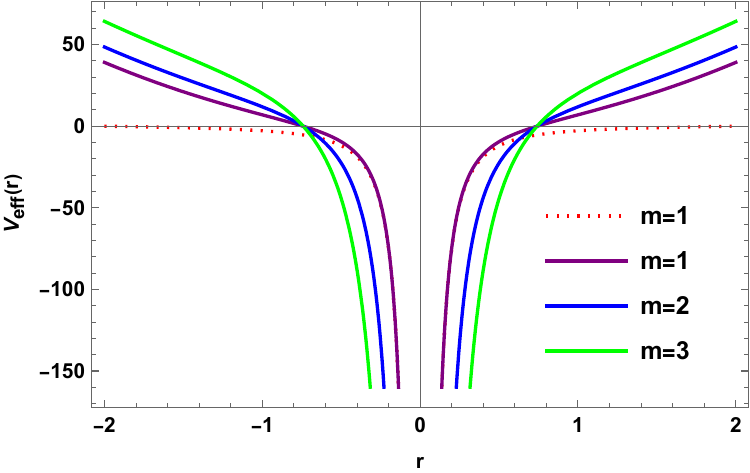}}\quad\quad
\subfloat[$\Lambda=-0.1$,$M=2$, $m=1$]{\centering{}\includegraphics[scale=0.65]{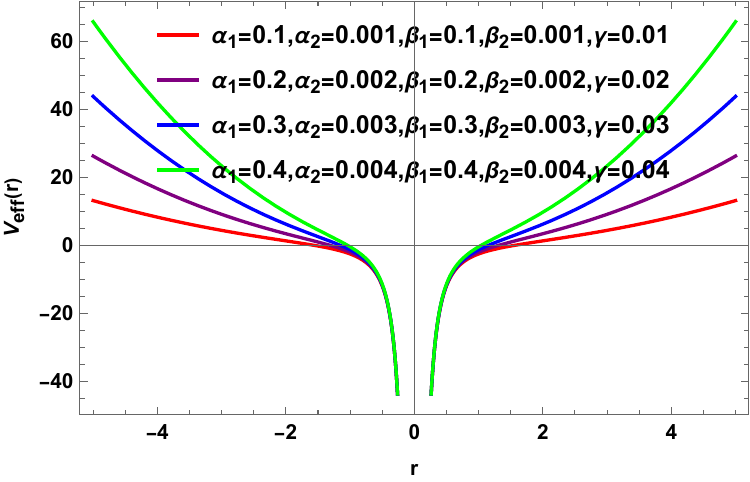}}
\hfill\\
\subfloat[$\Lambda=-0.1,M=-2$]{\centering{}\includegraphics[scale=0.65]{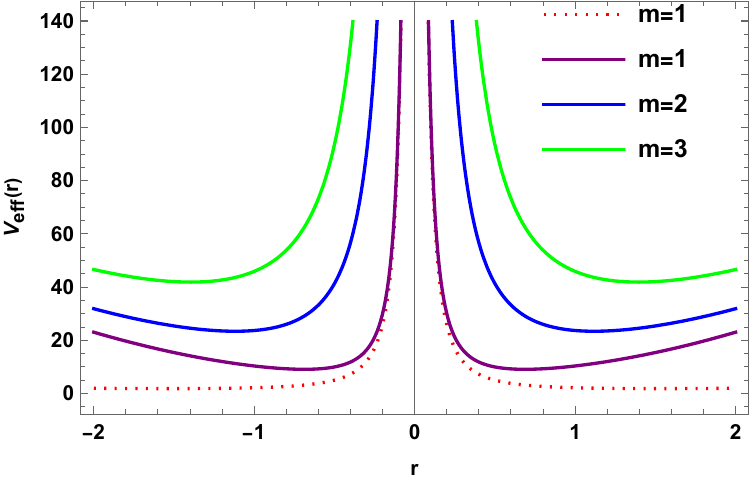}}\quad\quad
\subfloat[$\Lambda=-0.1$,$M=-2$, $m=1$]{\centering{}\includegraphics[scale=0.65]{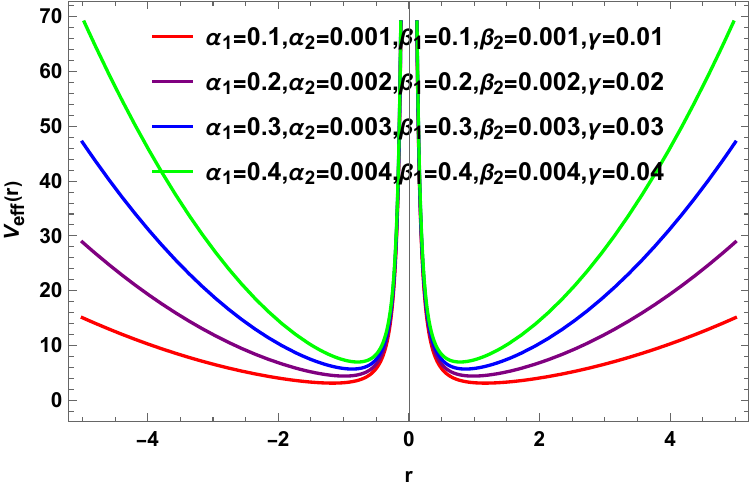}}
\centering{}\caption{The effective potential (\ref{EE4}) on non-rotating BTZ space-time in $\mathcal{RI}$-gravity.}\label{fig:5}
\end{figure}
\par\end{center}
The effect of variations of coupling constants $\alpha_i, \beta_i$ and $\gamma$ on the effective potential $V_{eff}$ on non-rotating BTZ space-time in $\mathcal{RI}$-gravity is illustrated in Figure \ref{fig:5}. The significant result on the effective potential is observed when the value of these coupling constants gradually increases for a particular $m$-state and compared with the GR result. In panel (a) of Figure \ref{fig:5}, the dotted red line representing $m=1$ corresponds to the GR case, where the parameters are set to $\alpha_i=0$, $\beta_i=0$, and $\gamma=0$. In contrast, the purple line for $m=1$ represents the $\mathcal{RI}$-gravity scenario, with the coupling constants set as $\alpha_1=0.8$, $\alpha_2=0.001$, $\beta_1=0.8$, $\beta_2=0.001$, and $\gamma=0.01$. The other lines in this panel depict different values of  $m$ while maintaining the same coupling constants. In panel (b) of Figure \ref{fig:5}, the lines correspond to various values of the coupling constants with $m=1$ state. These are the Figures where we set $M=2$. Similar explanation for Figure \ref{fig:5} (c)-(d) with $M=-2$.

\begin{center}
\begin{figure}[ht!]
\begin{centering}
\subfloat[$\Lambda=-0.1,M=2$]{\centering{}\includegraphics[scale=0.65]{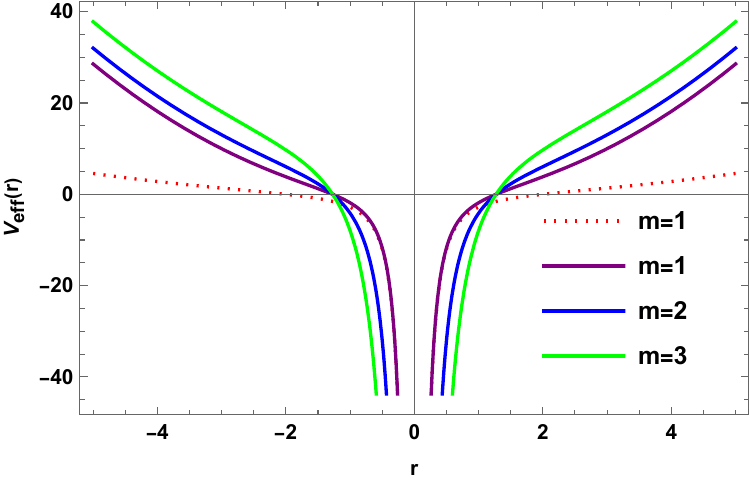}}\quad\quad
\subfloat[$\Lambda=-0.1$, $M=2$, $m=1$]{\centering{}\includegraphics[scale=0.65]{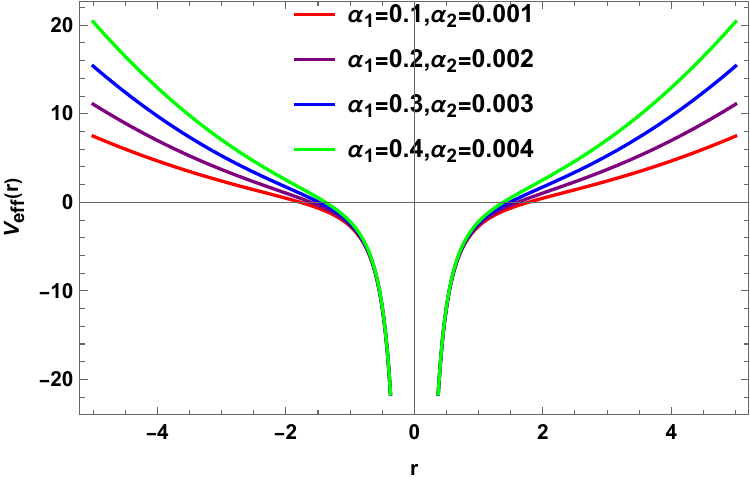}}
\hfill\\
\subfloat[$\Lambda=-0.1,M=-2$]{\centering{}\includegraphics[scale=0.65]{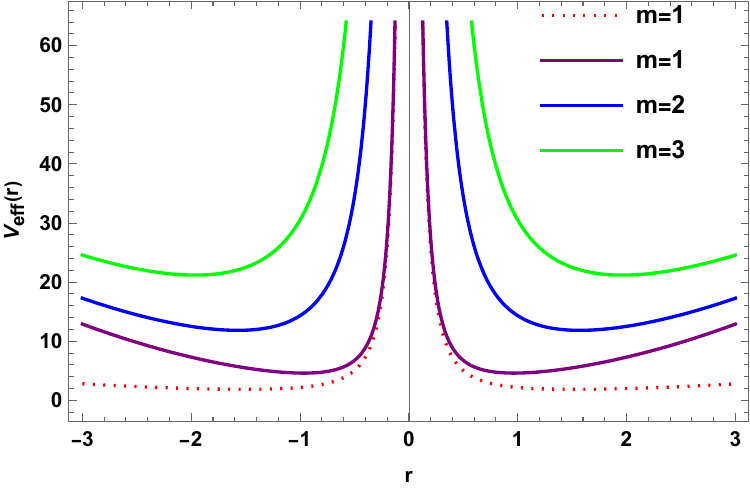}}\quad\quad
\subfloat[$\Lambda=-0.1$,$M=-2$, $m=1$]{\centering{}\includegraphics[scale=0.65]{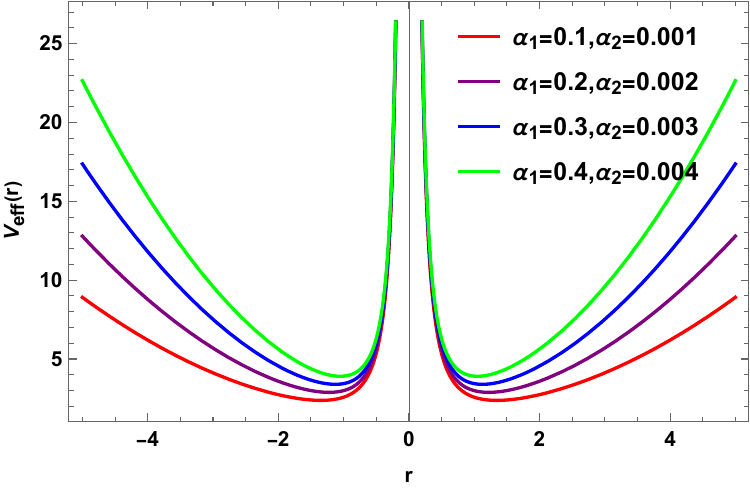}}
\centering{}\caption{The effective potential (\ref{EE4}) on non-rotating BTZ space-time in $f(\mathcal{R})$ gravity.}\label{fig:6}
\end{centering}
\end{figure}
\par\end{center}

The effect of variations of coupling constants $\alpha_i, \beta_i$ and $\gamma$ on the effective potential $V_{eff}$ on non-rotating BTZ space-time in $f(\mathcal{R})$-gravity is illustrated in Figure \ref{fig:6}. The significant result on the effective potential is observed when the value of these coupling constants gradually increases for a particular $m$-state and compared with the GR result. In panel (a) of Figure \ref{fig:3}, the dotted red line representing $m=1$ corresponds to the GR case, where the parameters are set to $\alpha_i=0$, $\beta_i=0$, and $\gamma=0$. In contrast, the purple line for $m=1$ represents the $RI$-gravity scenario, with the coupling constants set as $\alpha_1=0.5$, and $\alpha_2=0.001$. The other lines in this panel depict different values of  $m$ while maintaining the same coupling constants. In panel (b) of Figure \ref{fig:5}, the lines correspond to various values of the coupling constants with $m=1$ state. These are the figures where we set $M=2$. Similar scenario for Figure \ref{fig:6} (c)-(d) with $M=-2$.

\section{GUP-corrected thermodynamics of BTZ BH in modified gravity theories}\label{sec:3}

In recent years, the GUP has emerged as a promising extension of standard quantum mechanics, providing a phenomenological window into quantum gravity effects \cite{isTurakhonov:2024xfg,isChen:2024mlr,isAl-Badawi:2024cby,isSucu:2024gtr,isHoshimov:2023tlz,isSucu:2023moz,isSakalli:2022xrb,isBarman:2024hwd}. Its incorporation into black hole thermodynamics has revealed critical insights, especially for rotating black holes, which are key objects in understanding angular momentum in curved space-time. Here, we extend the application of GUP-modified thermodynamics to the rotating BTZ black hole within the framework of modified ($\mathcal{RI}$ or $f(\mathcal{R})$)-gravity (i.e., metrics (\ref{special-metric2}) or (\ref{ss6})). This approach not only allows us to analyze the thermodynamic properties influenced by the GUP but also opens up a broader understanding of modifications to Hawking radiation. Particularly, we derive corrections to the Hawking temperature and tunneling probability for massive bosons and fermions from the rotating BTZ black hole. We then analyze whther these GUP modifications yield intriguing implications for black hole evaporation or not. 

\subsection{Rotating BTZ black hole in dragging coordinates}\label{sec:31}

Before stating to the thermodynamical analysis, let us express the metric of rotating BTZ black hole of modified gravity theories in dragged coordinate system via the angular velocity at the horizon. To this end, we can re-express the line-element \eqref{a1} with metrics (\ref{special-metric2}) or (\ref{ss6}) as follows
\begin{equation} \label{is1}
ds^2 = h(r) \, dt^2 + \frac{1}{F(r)} \, dr^2 - J \, dt \, d\phi + L(r)^2 \, d\phi^2,
\end{equation}
where $h(r)$, $F(r)$, and $L(r)$ are functions of the radial coordinate $r$. By comparing metrics \eqref{a1} and \eqref{is1}, we get
\begin{equation}
h(r) = M + \Lambda_m r^2, \quad F(r) = -(M + \Lambda_m r^2)+ \frac{J^2}{4 r^2}, \quad L(r)^2 = r^2.
\end{equation}

The event horizon radius $r_+$ is determined by setting $F(r) = 0$, which gives:
\begin{equation}
r_+^2 = \frac{M}{2 \Lambda_m} \left( 1 + \sqrt{1 - \frac{J^2 \Lambda_m}{M^2}} \right).
\end{equation}

For small values of $J$, this can be approximated as:
\begin{equation}
r_+^2 \approx \frac{M}{2 \Lambda_m} + \frac{J^2}{8 M \Lambda_m}.
\end{equation}

The angular velocity $\Omega$ is calculated as:
\begin{equation} \label{isang}
\Omega =-\frac{g_{t\phi}}{g_{\phi\phi}}= \frac{J}{2 r^2},
\end{equation}
which takes the following form at the event horizon:
\begin{equation}
\Omega_H =\Omega\bigg|_{r=r_+} = \frac{J}{2 r_+^2}.
\end{equation}

To rewrite the metric in terms of the dragged coordinates, let us introduce:
\begin{equation}
d\phi = d\psi + \Omega dt.
\end{equation}

Substituting this transformation into the metric \eqref{is1} yields:
\begin{equation}\label{ismain}
ds^2 = G(r) \, dt^2 + \frac{1}{F(r)} \, dr^2 + L(r)^2 d\psi^2,
\end{equation}

in which
\begin{equation}
G(r)=  h(r) + \frac{J^2}{4 r^2}.
\end{equation}

\subsection{Massive Boson Emission from Rotating BTZ Metric in Modified Gravity Theories}\label{sec:32}

To analyze the quantum gravity effect on the tunneling process of a massive scalar particle from the rotating BTZ black hole in modified gravity theories, we use the modified Klein–Gordon equation \cite{isGecim:2017zid,isSakalli:2016mnk} in the presence of the GUP:
\begin{equation} \label{isKGE}
\hbar^2\, \partial_t \partial^t \mathcal{\Phi} + \hbar^2\, \partial_i \partial^i \mathcal{\Phi} + 2\, \alpha\, \hbar^4\, \partial_i \partial^i (\partial_i \partial^i \mathcal{\Phi}) + \mu_0^2\, (1 - 2\, \alpha\, \mu_0^2)\, \mathcal{\Phi} = 0,
\end{equation}
where $\alpha$ is the GUP parameter, $\hbar$ is the reduced Planck constant, $\mu_0$ is the mass of the scalar particle, and $\mathcal{\Phi}$ is the wave function.

Using the metric components of the dragged rotating BTZ black hole metric \eqref{ismain}, the Klein–Gordon equation \eqref{isKGE} becomes:
\begin{multline}\label{ismain2}
\hbar^2 \left( \frac{1}{g_{tt}} \frac{\partial^2 \Phi}{\partial t^2} +\frac{1}{g_{rr}} \frac{\partial^2 \Phi}{\partial r^2} + \frac{1}{g_{\psi \psi}} \frac{\partial^2 \mathcal{\Phi}}{\partial \psi^2} \right) + 2 \alpha \hbar^4 \frac{1}{g_{rr}} \frac{\partial^2}{\partial r^2} \left( \frac{1}{g_{rr}} \frac{\partial^2 \mathcal{\Phi}}{\partial r^2} \right) \\
+ \frac{1}{g_{\psi \psi}^2} \left(\frac{\partial^4 \mathcal{\Phi}}{\partial \psi^4} \right) + \mu_0^2 (1 - 2 \alpha \mu_0^2) \mathcal{\Phi} = 0.
\end{multline}

Following the WKB approximation \cite{isGecim:2019pft}, we can set the following ansatz:
\begin{equation} \label{isans1}
\mathcal{\Phi}(t, r, \psi) = \mathds{A} \,\exp \left( \frac{i}{\hbar}\, \mathcal{I}(t, r, \psi) \right),
\end{equation}
where $\mathds{A}$ is a slowly varying amplitude and $\mathcal{I}(t, r, \psi)$ is the classical action. Substituting Eq. \eqref{isans1} into Eq. \eqref{ismain2} and neglecting the higher-order terms in $\hbar$, one finds out the GUP-modified Hamilton–Jacobi equation:
\begin{multline}
\frac{1}{g_{tt}}\, \left( \frac{\partial \mathcal{I}}{\partial t} \right)^2 + \frac{1}{g_{rr}}\, \left( \frac{\partial \mathcal{I}}{\partial r} \right)^2 +\frac{1}{g_{\psi \psi}}\, \left( \frac{\partial \mathcal{I}}{\partial \psi} \right)^2 + 2\, \alpha\, \left[\frac{1}{g_{\psi \psi}^2} \left( \frac{\partial \mathcal{I}}{\partial \psi} \right)^4 + \frac{1}{g_{rr}^2} \left( \frac{\partial \mathcal{I}}{\partial r} \right)^4 \right] \\
- \mu_0^2 (1 - 2 \alpha \mu_0^2) = 0.
\end{multline}

Based on the symmetries of the metric \eqref{ismain}, one can assume a solution for the action $\mathcal{I}(t, r, \psi)$ of the form:
\begin{equation}
\mathcal{I}(t, r, \psi)=-(\mathcal{E}- j\,\Omega_H)\,t + j\,\psi + \mathrm{R}(r),
\end{equation}
where $\mathcal{E}$ and $j$ are the energy and angular momentum of the boson, respectively, and recall that $\Omega_H$ was the angular velocity at the horizon, see Eq.\eqref{isang}. The radial part $R(r)$ is obtained by integrating:
\begin{equation} \label{isR}
\mathrm{R}_{\pm}(r) = \pm \int  \left(1 + \alpha \Xi\right) \sqrt{\frac{(\mathcal{E} - j \Omega_H)^2 - g_{tt} \left(\mu_0^2 + \frac{j^2}{g_{\psi \psi}}\right)}{g_{tt} g_{rr}}} \,  dr,
\end{equation}
where:
\begin{equation}
\Xi = \frac{G^2(\mu_0^4 + j^4 / L^4) - [(E - j \Omega_H)^2 - G(\mu_0^2 - j^2 / L^2)]^2}{G[(E - j \Omega_H)^2 - G(\mu_0^2 - j^2 / L^2)]}.
\end{equation}

After integrating Eq. \eqref{isR} around the pole at the horizon, we find:
\begin{equation}
\mathrm{R}_{ \pm}\left({r}_{+}\right)= \pm i \pi \frac{l^2\left(E-j \Omega_{H}\right)\left(2 \tilde{r}_{+}^2\right)^{3 / 2}}{\Delta}\left[1+\alpha \mathcal{\S}_1\right],
\end{equation}
where
\begin{align} \label{isDel}
\Delta&=\sqrt{2}\left(4 r_{+}^4-\frac{J^2}{\Lambda_m} \right) , \\
\mathcal{\S}_1&=3\frac{\mu_0^2{r}_{+}^2+j^2}{2 {r}_{+}^2} .
\end{align}
The tunneling probabilities for outgoing and incoming particles \cite{isSakalli:2015jaa}
 are given by
\begin{equation}
\mathcal{P}_{\text{out}} = \exp \left( -\frac{2}{\hbar} \operatorname{Im} \mathrm{R}_+(r_+) \right), \quad \mathcal{P}_{\text{in}} = \exp \left( -\frac{2}{\hbar} \operatorname{Im} \mathrm{R}_-(r_+)\right).
\end{equation}

The total tunneling probability \cite{isAkhmedov:2006pg}  of the emitted bosons is given by
\begin{equation}
\Gamma_B = \frac{\mathcal{P}_{\text{out}}}{\mathcal{P}_{\text{in}}} = e^{-\frac{E}{T}},
\end{equation}
where $T$ is the temperature of the black hole. Whence, one can derive the GUP-modified Hawking temperature of the rotating BTZ black hole as follows
\begin{equation}
T_{HB}^{\text{GUP}} = \frac{T_H}{1 + \alpha \mathcal{\S}_1}\approx T_H(1 - \alpha \mathcal{\S}_1),
\end{equation}
where $T_H$ is the standard Hawking temperature \cite{isWald:1984rg}:
\begin{equation} \label{isTH}
T_H = \frac{\hbar\kappa}{2\pi}=\frac{\hbar \left(4 r_+^4 - J^2\right)}{4 \pi (2 r_+^2)^{3/2}}.
\end{equation}
In summary, the GUP-modified Hawking temperature of the rotating BTZ black hole of the modified gravity theories originating from the emitted bosons reads
\begin{equation}
T_{HB}^{\text{GUP}} = \frac{\hbar \left(4 r_+^4 - J^2\right)}{4 \pi (2 r_+^2)^{3/2}} \left(1 - 3\alpha \frac{j^2 + \mu_0^2 r_+^2}{4 r_+^2}\right),
\end{equation}
which can be explicitly written as follows
\begin{equation}
T_{HB}^{\text{GUP}} = \frac{\hbar \left( \frac{M^2}{\Lambda_m^2} \left( 1 + \sqrt{1 - \frac{J^2 \Lambda_m}{M^2}} \right)^2 - J^2 \right)}{4 \pi \left( \frac{M}{\Lambda_m} \left( 1 + \sqrt{1 - \frac{J^2 \Lambda_m}{M^2}} \right) \right)^{3/2}} \left( 1 - \frac{3 \alpha \left( j^2 + \mu_0^2 \frac{M}{2 \Lambda_m} \left( 1 + \sqrt{1 - \frac{J^2 \Lambda_m}{M^2}} \right) \right)}{2 M \left( 1 + \sqrt{1 - \frac{J^2 \Lambda_m}{M^2}} \right)} \right).
\end{equation}

In the absence of GUP effect, $\alpha=0$, the Hawking temperature reduces to
\begin{eqnarray}
    T_{HB}= \frac{\hbar \left( \frac{M^2}{\Lambda_m^2} \left(1+\sqrt{1-\frac{J^2\,\Lambda_m}{M^2}}\right)^2- J^2 \right)}{4\,\pi\,\left( \frac{M}{\Lambda_m} \left(1 + \sqrt{1-\frac{J^2\,\Lambda_m}{M^2}}\right) \right)^{3/2}}.\label{KK1} 
\end{eqnarray}

In the context of $\mathcal{RI}$ gravity, the Hawking temperature from Eq.~(\ref{KK1}) is derived as follows:

\small
\begin{eqnarray}
    T_{HB}^{\mathcal{RI}} &=& \frac{\hbar/4\pi}{\left[\frac{M}{\Lambda - 6\,\alpha_1\,\Lambda^2 - 108\,\alpha_2\,\Lambda^3 + \frac{5\,\beta_1}{4\,\Lambda} + \frac{15\,\beta_2}{8\,\Lambda^2} + \frac{7\,\gamma}{8\,\Lambda^2}} \left(1 + \sqrt{1 - \frac{J^2 \left(\Lambda - 6\,\alpha_1\,\Lambda^2 - 108\,\alpha_2\,\Lambda^3 + \frac{5\,\beta_1}{4\,\Lambda} + \frac{15\,\beta_2}{8\,\Lambda^2} + \frac{7\,\gamma}{8\,\Lambda^2}\right)}{M^2}}\right) \right]^{3/2}} \nonumber \\
    && \times \Bigg[\frac{M^2}{\left(\Lambda - 6\alpha_1\Lambda^2 - 108\alpha_2\Lambda^3 + \frac{5\beta_1}{4\Lambda} + \frac{15\beta_2}{8\Lambda^2} + \frac{7\gamma}{8\Lambda^2}\right)^2} \Bigg\{1 + \sqrt{1 - \frac{J^2}{M^2} \left(\Lambda - 6\alpha_1\Lambda^2 - 108\alpha_2\Lambda^3 + \frac{5\beta_1}{4\Lambda} + \frac{15\beta_2}{8\Lambda^2} + \frac{7\gamma}{8\Lambda^2}\right)}\Bigg\}^2 \nonumber \\
    && - J^2\Bigg].\label{KK2} 
\end{eqnarray}
\normalsize

The above expression shows that the Hawking temperature is influenced by the coupling constants $\alpha_i$, $\beta_i$, and $\gamma$ within the framework of $\mathcal{RI}$ gravity theory, leading to modifications relative to the case of GR.

For the $f(\mathcal{R})$-gravity theory, the Hawking temperature from Eq.~(\ref{KK1}) is derived as follows:
\small
\begin{eqnarray}
    T_{HB}^{f(\mathcal{R})} &=& \frac{\hbar/4\pi}{\left[\frac{M}{\Lambda - 6\,\alpha_1\,\Lambda^2 - 108\,\alpha_2\,\Lambda^3} \left(1 + \sqrt{1 - \frac{J^2 \left(\Lambda - 6\,\alpha_1\,\Lambda^2 - 108\,\alpha_2\,\Lambda^3\right)}{M^2}}\right) \right]^{3/2}} \nonumber \\
    && \times \Bigg[\frac{M^2}{\left(\Lambda - 6\alpha_1\Lambda^2 - 108\alpha_2\Lambda^3\right)^2} \Bigg\{1 + \sqrt{1 - \frac{J^2}{M^2} \left(\Lambda - 6\alpha_1\Lambda^2 - 108\alpha_2\Lambda^3\right)}\Bigg\}^2 - J^2\Bigg].\label{KK3} 
\end{eqnarray}
\normalsize

In this case, the Hawking temperature is influenced by the coupling constants $\alpha_i$ in the context of $f(\mathcal{R})$ gravity, showing further deviations from GR.

When accounting for the GUP effect, the Hawking temperature for the BTZ black hole in $\mathcal{RI}$ gravity is modified to:
\small
\begin{eqnarray}
     T_{HB}^{\mathcal{RI}} &=& \frac{\hbar/4\pi}{\left[\frac{M}{\Lambda - 6\,\alpha_1\,\Lambda^2 - 108\,\alpha_2\,\Lambda^3 + \frac{5\,\beta_1}{4\,\Lambda} + \frac{15\,\beta_2}{8\,\Lambda^2} + \frac{7\,\gamma}{8\,\Lambda^2}} \left(1 + \sqrt{1 - \frac{J^2 \left(\Lambda - 6\,\alpha_1\,\Lambda^2 - 108\,\alpha_2\,\Lambda^3 + \frac{5\,\beta_1}{4\,\Lambda} + \frac{15\,\beta_2}{8\,\Lambda^2} + \frac{7\,\gamma}{8\,\Lambda^2}\right)}{M^2}}\right) \right]^{3/2}} \nonumber \\
    && \times \Bigg[\frac{M^2}{\left(\Lambda - 6\alpha_1\Lambda^2 - 108\alpha_2\Lambda^3 + \frac{5\beta_1}{4\Lambda} + \frac{15\beta_2}{8\Lambda^2} + \frac{7\gamma}{8\Lambda^2}\right)^2} \Bigg\{1 + \sqrt{1 - \frac{J^2}{M^2} \left(\Lambda - 6\alpha_1\Lambda^2 - 108\alpha_2\Lambda^3 + \frac{5\beta_1}{4\Lambda} + \frac{15\beta_2}{8\Lambda^2} + \frac{7\gamma}{8\Lambda^2}\right)}\Bigg\}^2 \nonumber \\
    && - J^2\Bigg] \, \Bigg[1 - \frac{3\,\alpha}{2\,M} \Bigg\{\frac{j^2}{1 + \sqrt{1 - \frac{J^2 \left(\Lambda - 6\,\alpha_1\,\Lambda^2 - 108\,\alpha_2\,\Lambda^3 + \frac{5\,\beta_1}{4\,\Lambda} + \frac{15\,\beta_2}{8\,\Lambda^2} + \frac{7\,\gamma}{8\,\Lambda^2}\right)}{M^2}}} \nonumber \\
    && + \frac{\mu^2_{0}\,M/2}{\left(\Lambda - 6\,\alpha_1\,\Lambda^2 - 108\,\alpha_2\,\Lambda^3 + \frac{5\,\beta_1}{4\,\Lambda} + \frac{15\,\beta_2}{8\,\Lambda^2} + \frac{7\,\gamma}{8\,\Lambda^2}\right)} \Bigg\}\Bigg].\label{KK4} 
\end{eqnarray}
\normalsize

In the presence of the GUP effect, the Hawking temperature is influenced by the coupling constants $\alpha_i$, $\beta_i$, and $\gamma$ in $RI$-gravity, showing additional modifications compared to GR.

For $f(\mathcal{R})$ gravity, the Hawking temperature under the GUP effect is:
\small
\begin{eqnarray}
    T_{HB}^{f(\mathcal{R})} &=& \frac{\hbar/4\pi}{\left[\frac{M}{\Lambda - 6\,\alpha_1\,\Lambda^2 - 108\,\alpha_2\,\Lambda^3} \left(1 + \sqrt{1 - \frac{J^2 \left(\Lambda - 6\,\alpha_1\,\Lambda^2 - 108\,\alpha_2\,\Lambda^3\right)}{M^2}}\right) \right]^{3/2}} \nonumber \\
    && \times \Bigg[\frac{M^2}{\left(\Lambda - 6\alpha_1\Lambda^2 - 108\alpha_2\Lambda^3\right)^2} \Bigg\{1 + \sqrt{1 - \frac{J^2}{M^2} \left(\Lambda - 6\alpha_1\Lambda^2 - 108\alpha_2\Lambda^3\right)}\Bigg\}^2 - J^2\Bigg] \nonumber \\
    && \times \Bigg[1 - \frac{3\,\alpha}{2\,M} \Bigg\{\frac{j^2}{1 + \sqrt{1 - \frac{J^2 \left(\Lambda - 6\,\alpha_1\,\Lambda^2 - 108\,\alpha_2\,\Lambda^3\right)}{M^2}}} + \frac{\mu^2_{0}\,M/2}{\left(\Lambda - 6\,\alpha_1\,\Lambda^2 - 108\,\alpha_2\,\Lambda^3\right)} \Bigg\}\Bigg].\label{KK5} 
\end{eqnarray}
\normalsize

In $f(\mathcal{R})$-gravity framework with the GUP effect, the Hawking temperature depends on the coupling constants $\alpha_i$, resulting in further deviations from the GR case.

\subsection{Emission of Massive Fermions from a Rotating BTZ Metric in Modified Gravity Theories}\label{sec:33}

In this section, we examine the emission of a massive Dirac particles from the rotating BTZ metric of the modified gravity theories, taking into account the influence of the GUP on the metric \eqref{ismain}. The GUP-modified Dirac equation is as follows \cite{isGecim:2018sji}: 
\begin{align} \label{isDrc}
i {\sigma}^0(q) \partial_0 \Psi_D + i {\sigma}^i(q) (1 - \alpha \mu_D^2) \partial_i \Psi_D - \frac{\mu_D}{\hbar} \left( 1 + \alpha \hbar^2 \partial_i \partial^i - \alpha \mu_D^2 \right) \Psi_D \nonumber \\
+ i \alpha \hbar^2 {\sigma}^i(q) \partial_i (\partial^j \partial_j \Psi_D) - i {\sigma}^\mu(q) \Gamma_\mu(q) \Psi_D \left( 1 + \alpha \hbar^2 \partial_i \partial^i - \alpha \mu_D^2 \right) = 0,
\end{align}
where $\Psi_D$ is the modified Dirac spinor, $\mu_D$ is the mass of the Dirac particle, ${\sigma}^\mu(q)$ are the space-time-dependent Dirac matrices, and $\Gamma_\mu(q)$ represents the spin affine connection. Here, $q$ is a generalized coordinate that represents $(t,r, \psi)$. The spin connection coefficients $\Gamma_\mu(q)$, needed for the curved space-time Dirac equation, are expressed in terms of the vielbein fields and the Christoffel symbols:
\begin{equation}
\Gamma_\mu(q) = \frac{1}{8} g_{\nu \alpha} \left( e_i^{\ \nu} \partial_\mu e^{\alpha}_j - \Gamma^\alpha_{\mu \nu} e^\nu_j \right) [{\sigma}^i(q), {\sigma}^j(q)].
\end{equation}
In the dragged metric \eqref{ismain}, the non-zero spin connection terms are:
\begin{equation}
\Gamma_0 = -\frac{i}{4} \sqrt{\frac{F}{h}} \frac{dG}{dr} \sigma^3 \sigma^1, \quad \Gamma_2 = \frac{1}{2} \sqrt{\frac{F}{L}} \frac{dL}{dr} \sigma^1 \sigma^2.
\end{equation}

Emplotying the WKB approximation, we write the modified Dirac spinor as:
\begin{equation}
\Psi_D(q) = \exp \left( \frac{i}{\hbar} \mathcal{I}(q) \right) \begin{pmatrix} \mathrm{A}_+(q) \\ \mathrm{B}_+(q) \end{pmatrix},
\end{equation}
where $\mathrm{A}_+(q)$ and $\mathrm{B}_+(q)$ represent the spin-$\uparrow$ (positive energy) and spin-$\downarrow$ (negative energy) components, respectively. Substituting the spin connection terms, the WKB approximation, and the spinor ansatz into the GUP-modified Dirac equation \eqref{isDrc}, we obtain the following reduced form of the GUP-modified Dirac equation for the metric \eqref{ismain}:

\begin{eqnarray} \label{isDirac}
&\mathrm{B}\left[ i \sqrt{F(r)} (1 - \alpha \mu_D^2) \left( \frac{\partial \mathcal{I}}{\partial r} \right) + \frac{(1 - \alpha \mu_D^2)}{L} \left( \frac{\partial \mathcal{I}}{\partial \psi} \right) + i \alpha \sqrt{F(r)} \left( \frac{\partial \mathcal{I}}{\partial r} \right) \left( \frac{\partial \mathcal{I}}{\partial \psi} \right)^2 \right] \nonumber \\
& + \mathrm{B}\left[ \frac{\alpha F(r)}{L} \left( \frac{\partial \mathcal{I}}{\partial \psi} \right) \left( \frac{\partial \mathcal{I}}{\partial r} \right)^2 + i \alpha F(r)^{3/2} \left( \frac{\partial \mathcal{I}}{\partial r} \right)^3 + \frac{\alpha}{L^3} \left( \frac{\partial \mathcal{I}}{\partial \psi} \right)^3 \right] \nonumber \\
& + \mathrm{A} \left[ \frac{1}{\sqrt{G(r)}} \left( \frac{\partial \mathcal{I}}{\partial t} \right) + \mu_D (1 - \alpha \mu_D^2) + \frac{\alpha \mu_D}{L^2} \left( \frac{\partial \mathcal{I}}{\partial \psi} \right)^2 + \alpha \mu_D F(r) \left( \frac{\partial \mathcal{I}}{\partial r} \right)^2 \right] = 0, \\
&\mathrm{A}\left[ -i \sqrt{F(r)} (1 - \alpha \mu_D^2) \left( \frac{\partial \mathcal{I}}{\partial r} \right) + \frac{(1 - \alpha \mu_D^2)}{L} \left( \frac{\partial \mathcal{I}}{\partial \psi} \right) - i \alpha \sqrt{F(r)} \left( \frac{\partial I}{\partial r} \right) \left( \frac{\partial \mathcal{I}}{\partial \psi} \right)^2 \right] \nonumber \\
& + \mathrm{A} \left[ \frac{\alpha}{L^3} \left( \frac{\partial \mathcal{I}}{\partial \psi} \right)^3 - i \alpha F(r)^{3/2} \left( \frac{\partial \mathcal{I}}{\partial r} \right)^3 + \frac{F(r)}{L} \left( \frac{\partial \mathcal{I}}{\partial \psi} \right) \left( \frac{\partial \mathcal{I}}{\partial r} \right)^2 \right] \nonumber \\
& + \mathrm{B} \left[ \frac{1}{\sqrt{G(r)}} \left( \frac{\partial I}{\partial t} \right) - \mu_D \left( 1 - \alpha \mu_D^2 \right) - \frac{\alpha \mu_D}{L^2} \left( \frac{\partial \mathcal{I}}{\partial \psi} \right)^2 - \alpha \mu_D F(r) \left( \frac{\partial \mathcal{I}}{\partial r} \right)^2 \right] = 0. \label{isDirac2}
\end{eqnarray}

We proceed by taking the action as
\begin{equation} \label{isans2}
\mathcal{I}(t, r, \phi) = -(\omega - k \Omega_H) t + k \phi + \mathrm{R}(r),
\end{equation}
where $\omega$ and $k$ are the fermion's energy and angular momentum, respectively. The non-trivial solution for Eqs. \eqref{isDirac} and \eqref{isDirac2} emerges when the determinant of the coefficient matrix equals zero, expressed as $\mathds{M}(A, B)^T=0$. Ignoring terms of higher order in the $\alpha$ parameter gives rise to the modified Hamilton-Jacobi equation applicable to a massive Dirac particle:
\begin{align}\label{isMDirac}
\frac{1}{G(r)} \left( \frac{\partial \mathcal{I}}{\partial t} \right)^2 - F(r) \left( \frac{\partial \mathcal{I}}{\partial r} \right)^2 - \frac{1}{L(r)^2} \left( \frac{\partial \mathcal{I}}{\partial \phi} \right)^2 - \mu_D^2 \left( 1 - 2 \alpha \mu_D^2 \right) \nonumber \\
- \alpha \frac{4 F(r)}{L(r)^2} \left( \frac{\partial \mathcal{I}}{\partial r} \right)^2 \left( \frac{\partial \mathcal{I}}{\partial \phi} \right)^2 - 2 \alpha \left[ \frac{1}{L(r)^4} \left( \frac{\partial \mathcal{I}}{\partial \phi} \right)^4 + F(r)^2 \left( \frac{\partial \mathcal{I}}{\partial r} \right)^4 \right] = 0.
\end{align}

By setting the appropriate substitutions from Eq. \eqref{isans2} into Eq. \eqref{isMDirac}, and solving for the radial trajectory $R_{\pm}(r)$, we get:
\begin{equation}
\mathrm{R}_{\pm}(r) = \pm \int \frac{\sqrt{(\omega - k \Omega_H)^2 - G(r) \left( \mu_D^2 + \frac{k^2}{L(r)^2} \right)}}{\sqrt{F(r) G(r)}} \left( 1 + \alpha \Xi \right) \, dr,
\end{equation}
where $\Xi$ is the GUP correction term defined by:
\begin{equation}
\Xi = \frac{(\omega - k \Omega_H)^2 \left( 2 \mu_D^2 G(r) - (\omega - k \Omega_H)^2 \right)}{G(r) \left( (\omega - k \Omega_H)^2 - G(r) \left( \mu_D^2 + \frac{k^2}{L(r)^2} \right) \right)}.
\end{equation}

To calculate the imaginary part of the action near the horizon, we integrate around the pole at $r = r_+$:
\begin{equation}
\operatorname{Im} \, S = \operatorname{Im} \int_{r_+}^r \frac{\sqrt{(\omega - k \Omega_H)^2 - G(r) \left( \mu_D^2 + \frac{k^2}{L(r)^2} \right)}}{\sqrt{F(r) G(r)}} \left( 1 + \alpha \Xi \right) \, dr.
\end{equation}

Calculating this integral gives:
\begin{equation}
\mathrm{R}_{\pm}(r_+) = \pm i \pi \frac{(\omega - k \Omega_H) \left( 2 r_+^2 \right)^{3/2}}{\Delta} \left( 1 + \alpha \mathcal{\S}_2 \right),
\end{equation}
where $\Delta$ is a function of black hole parameters given in Eq. \eqref{isDel}, and $\mathcal{\S}_2$ represents the GUP corrections. Following the previous subsection, one can obtain the tunneling probability for the Dirac particle emission as
\begin{equation}
\Gamma_D \sim e^{-2 \operatorname{Im} \mathrm{R}_+ (r_+)}.
\end{equation}

As is well-known, from the tunneling rate, we can identify the GUP-modified Hawking temperature $T_{HD}^{\text{GUP}}$ of the emitted fermions from the rotating BTZ black hole as follows:
\begin{equation} \label{isTHD}
T_{HD}^{\text{GUP}} = T_H \left( 1 - \alpha \mathcal{\S}_2 \right),
\end{equation}
where $T_H$ is the standard Hawking temperature \eqref{isTH} for the rotating BTZ metric of the modified gravity theories, and $\mathcal{\S}_2$ is given by
\begin{equation}
\mathcal{\S}_2 = \frac{3 \mu_D^2 r_+^2 - k^2}{2 r_+^2}.
\end{equation}
The modified Hawking temperature \eqref{isTHD} can be explicitly written as follows
\begin{equation}\label{Haw}
T_{HD}^{\text{GUP}} = \frac{\hbar \left( \frac{M^2}{\Lambda_m^2} \left( 1 + \sqrt{1 - \frac{J^2 \Lambda_m}{M^2}} \right)^2 - J^2 \right)}{4 \pi \left( \frac{M}{\Lambda_m} \left( 1 + \sqrt{1 - \frac{J^2 \Lambda_m}{M^2}} \right) \right)^{3/2}} \left( 1 - \frac{3 \alpha \left( \mu_D^2 \frac{M}{2 \Lambda_m} \left( 1 + \sqrt{1 - \frac{J^2 \Lambda_m}{M^2}} \right) - k^2 \right)}{\frac{M}{\Lambda_m} \left( 1 + \sqrt{1 - \frac{J^2 \Lambda_m}{M^2}} \right)} \right).
\end{equation}

In the absence of the GUP effect ($\alpha=0$), the Hawking temperature given in Eq.~(\ref{Haw}) simplifies to
\begin{eqnarray}
    T_{HD}= \frac{(\hbar/4\pi) \left[\frac{M^2}{\Lambda_m^2} \left(1+\sqrt{1-\frac{J^2\,\Lambda_m}{M^2}}\right)^2- J^2 \right]}{\left[\frac{M}{\Lambda_m} \left(1 + \sqrt{1-\frac{J^2\,\Lambda_m}{M^2}}\right) \right]^{3/2}}.\label{MM1} 
\end{eqnarray}
Within the frameworks of$\mathcal{RI}$-gravity and $f(\mathcal{R})$-gravity theories, the Hawking temperatures are given in Eqs.~(\ref{KK2}) and (\ref{KK3}), respectively. When the GUP effect is considered, however, the expression for the Hawking temperature is modified accordingly.

Taking the GUP effect into account, the Hawking temperature of the BTZ black hole within the context of $RI$-gravity becomes:
\small
\begin{eqnarray}
     T_{HD}^{\mathcal{RI}}&=&\frac{\hbar/4\pi}{\left[\frac{M}{\Lambda-6\,\alpha_1\,\Lambda^2-108\,\alpha_2\,\Lambda^3+\frac{5\,\beta_1}{4\,\Lambda}+\frac{15\,\beta_2}{8\,\Lambda^2}+\frac{7\,\gamma}{8\,\Lambda^2}} \left(1 + \sqrt{1-\frac{J^2\,\left(\Lambda-6\,\alpha_1\,\Lambda^2-108\,\alpha_2\,\Lambda^3+\frac{5\,\beta_1}{4\,\Lambda}+\frac{15\,\beta_2}{8\,\Lambda^2}+\frac{7\,\gamma}{8\,\Lambda^2}\right)}{M^2}}\right) \right]^{3/2}} \nonumber\\
    &&\times\Bigg[\frac{M^2}{\left(\Lambda-6\alpha_1\Lambda^2-108\alpha_2\Lambda^3+\frac{5\beta_1}{4\Lambda}+\frac{15\beta_2}{8\Lambda^2}+\frac{7\gamma}{8\Lambda^2}\right)^2} \Bigg\{1+\sqrt{1-\frac{J^2}{M^2}\left(\Lambda-6\alpha_1\Lambda^2-108\alpha_2\Lambda^3+\frac{5\beta_1}{4\Lambda}+\frac{15\beta_2}{8\Lambda^2}+\frac{7\gamma}{8\Lambda^2}\right)}\Bigg\}^2\nonumber\\
    &&-J^2\Bigg] \times \Bigg[1-3\,\alpha\,
    \Bigg\{\frac{\mu^2_{D}}{2}-\frac{(k^2/M)\,\left(\Lambda-6\,\alpha_1\,\Lambda^2-108\,\alpha_2\,\Lambda^3+\frac{5\,\beta_1}{4\,\Lambda}+\frac{15\,\beta_2}{8\,\Lambda^2}+\frac{7\,\gamma}{8\,\Lambda^2}\right)}{1+\sqrt{1-\frac{J^2\,\left(\Lambda-6\,\alpha_1\,\Lambda^2-108\,\alpha_2\,\Lambda^3+\frac{5\,\beta_1}{4\,\Lambda}+\frac{15\,\beta_2}{8\,\Lambda^2}+\frac{7\,\gamma}{8\,\Lambda^2}\right)}{M^2}}}
    \Bigg\}\Bigg].\label{MM2} 
\end{eqnarray}
\normalsize

The presence of the GUP effect significantly influences the Hawking temperature, as seen in the additional modifications imparted by the coupling constants $\alpha_i$, $\beta_i$, and $\gamma$ in $RI$-gravity theory. This effect leads to more complex dependencies compared to the GR case.

In the context of $f(\mathcal{R})$ gravity, the Hawking temperature is given by:
\small
\begin{eqnarray}
     T_{HD}^{f(\mathcal{R})}&=&\frac{\hbar/4\pi}{\left[\frac{M}{\Lambda-6\,\alpha_1\,\Lambda^2-108\,\alpha_2\,\Lambda^3} \left(1 + \sqrt{1-\frac{J^2\,\left(\Lambda-6\,\alpha_1\,\Lambda^2-108\,\alpha_2\,\Lambda^3\right)}{M^2}}\right) \right]^{3/2}} \nonumber\\
    &&\times\Bigg[\frac{M^2}{\left(\Lambda-6\alpha_1\Lambda^2-108\alpha_2\Lambda^3\right)^2} \Bigg\{1+\sqrt{1-\frac{J^2}{M^2}\left(\Lambda-6\alpha_1\Lambda^2-108\alpha_2\Lambda^3\right)}\Bigg\}^2
    -J^2\Bigg] \nonumber\\
    &&\times \Bigg[1-3\,\alpha\,
    \Bigg\{\frac{\mu^2_{D}}{2}-\frac{(k^2/M)\,\left(\Lambda-6\,\alpha_1\,\Lambda^2-108\,\alpha_2\,\Lambda^3\right)}{1+\sqrt{1-\frac{J^2\,\left(\Lambda-6\,\alpha_1\,\Lambda^2-108\,\alpha_2\,\Lambda^3\right)}{M^2}}}
    \Bigg\}\Bigg].\label{MM3} 
\end{eqnarray}
\normalsize

With the inclusion of the GUP effect in $f(\mathcal{R})$ gravity theory, the Hawking temperature is altered due to the influence of the coupling constants $\alpha_i$, leading to distinct modifications compared to GR.

In summary, when considering the GUP effect, the Hawking temperature of the BTZ black hole is affected by the coupling constants $\alpha_i$, $\beta_i$, and $\gamma$ within the $\mathcal{RI}$-gravity framework, as well as by $\alpha_i$ in the $f(\mathcal{R})$-gravity framework. This effect provides further insights into the behavior of massive bosons and fermions in these modified gravity theories.

\section{Geodesics of non-rotating BTZ space-time in modified gravity theories} \label{sec:4}

The equations governing the geodesics in the space-time given in (\ref{special-metric7}) can be
derived from the Lagrangian equation,
\begin{equation}
\frac{d}{ds}\left( \frac{\partial \mathcal{L}}{\partial \overset{\cdot }{%
x^{\mu }}}\right) =\frac{\partial \mathcal{L}}{\partial x^{\mu }},
\end{equation}%
where $s$ is the affine parameter of the light trajectory, a dot denotes a
differentiation with respect to $s$. 

The Lagrangian for the space-time described in (\ref{special-metric7}) is given by 
\begin{equation}
\mathcal{L}=\frac{1}{2}\left[ g_{tt}\dot{t}^{2}+ g_{rr}\dot{r}^{2}
+r^{2} \dot{\psi}^{2}\right] .
\end{equation}%

space-time metric tensor $g_{\mu \nu}$ depends only on $r$ and is independent of $t$ and $\psi$. Thus, the black hole has two Killing vectors. Those two constants of
motion which can be labeled as $E$ and $L$, are given by 
\begin{equation*}
E=-g_{tt}\dot{t}=-\left( M+\Lambda _{m}r^{2}\right) \dot{t},  \label{en1}
\end{equation*}%
\begin{equation}
L=r^{2}\dot{\psi}\text{.}  \label{ang1}
\end{equation}%
Using two constants of motion given in Eq. (\ref{ang1})
the geodesics equation becomes 
\begin{equation}
\left( \frac{dr}{d\phi }\right) ^{2}=\frac{r^{4}}{L^{2}}\left( E^{2}-\left(
M+\Lambda _{m}r^{2}\right) \right) \left( \epsilon +\frac{L^{2}}{r^{2}}%
\right) .
\end{equation}%

\begin{figure}[htbp]
 \includegraphics[width=0.48\linewidth]{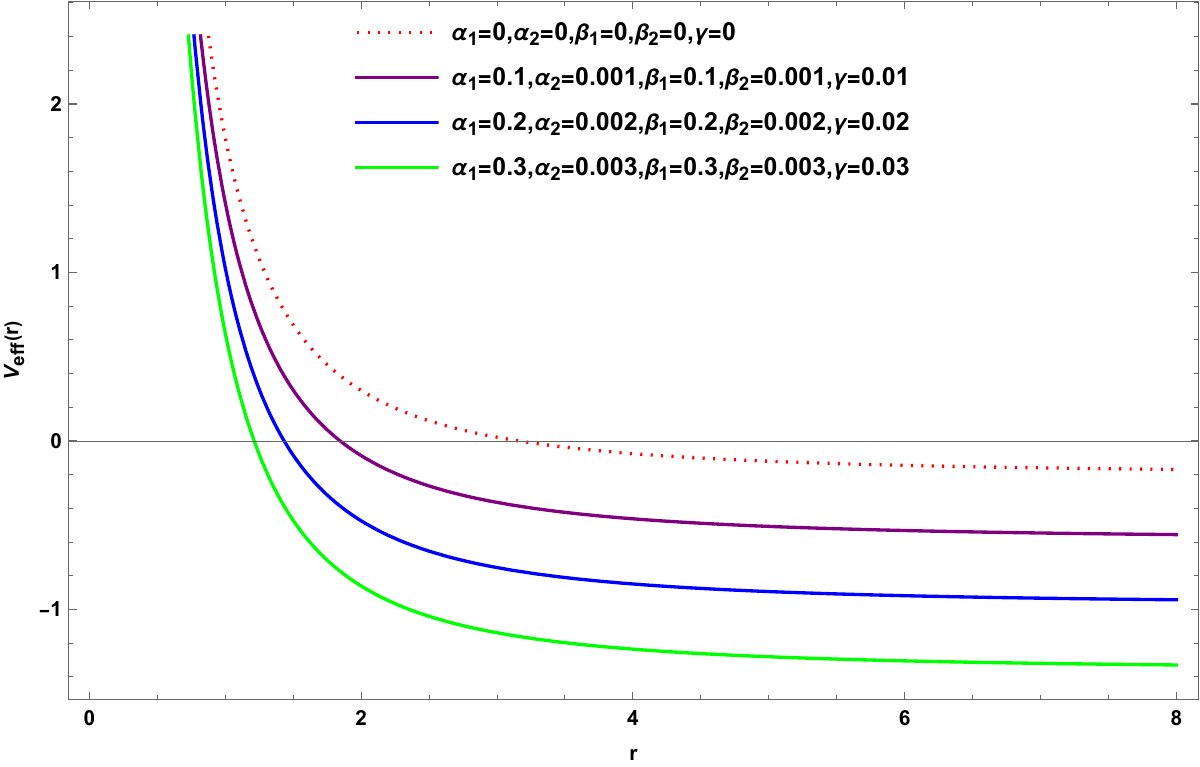}\quad\quad
 \includegraphics[width=0.48\linewidth]{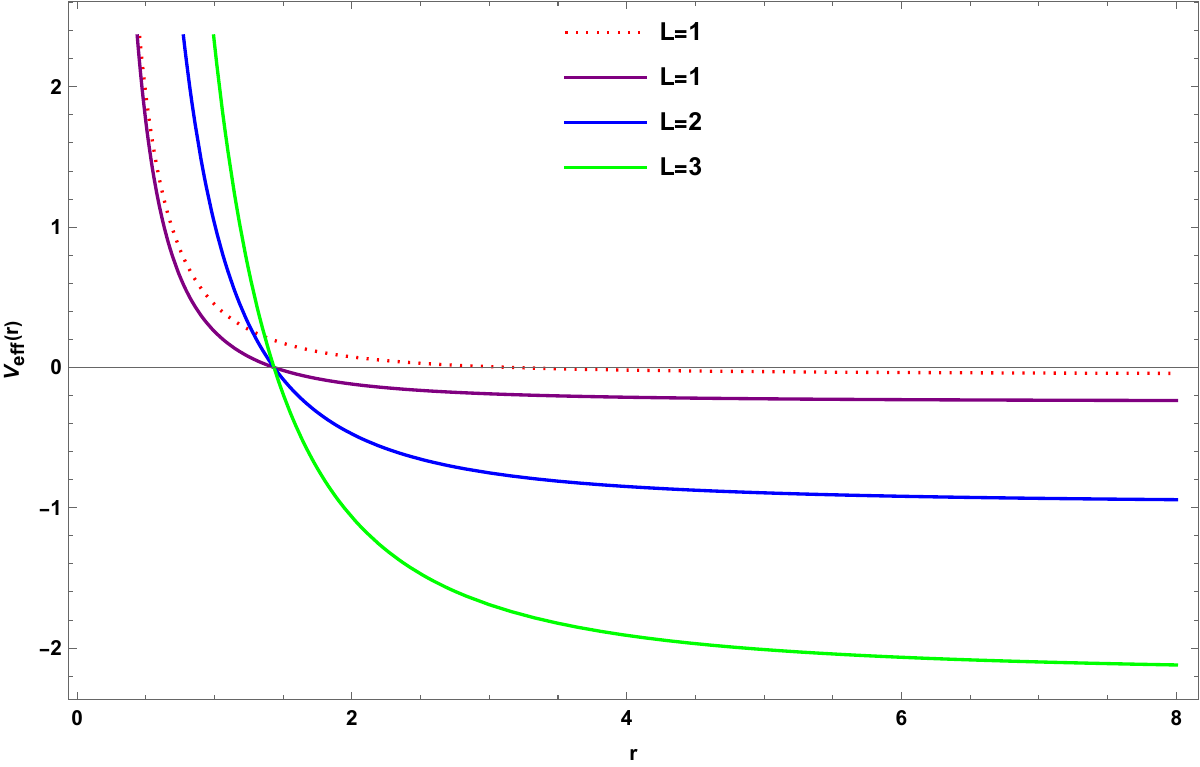}
    \caption{The potential for null geodesics in static BTZ space-time in $RI$-gravity paradigm. Here, $\Lambda=-0.1$ and $M=1$.}
    \label{fig:7}
\hfill\\
   \includegraphics[width=0.48\linewidth]{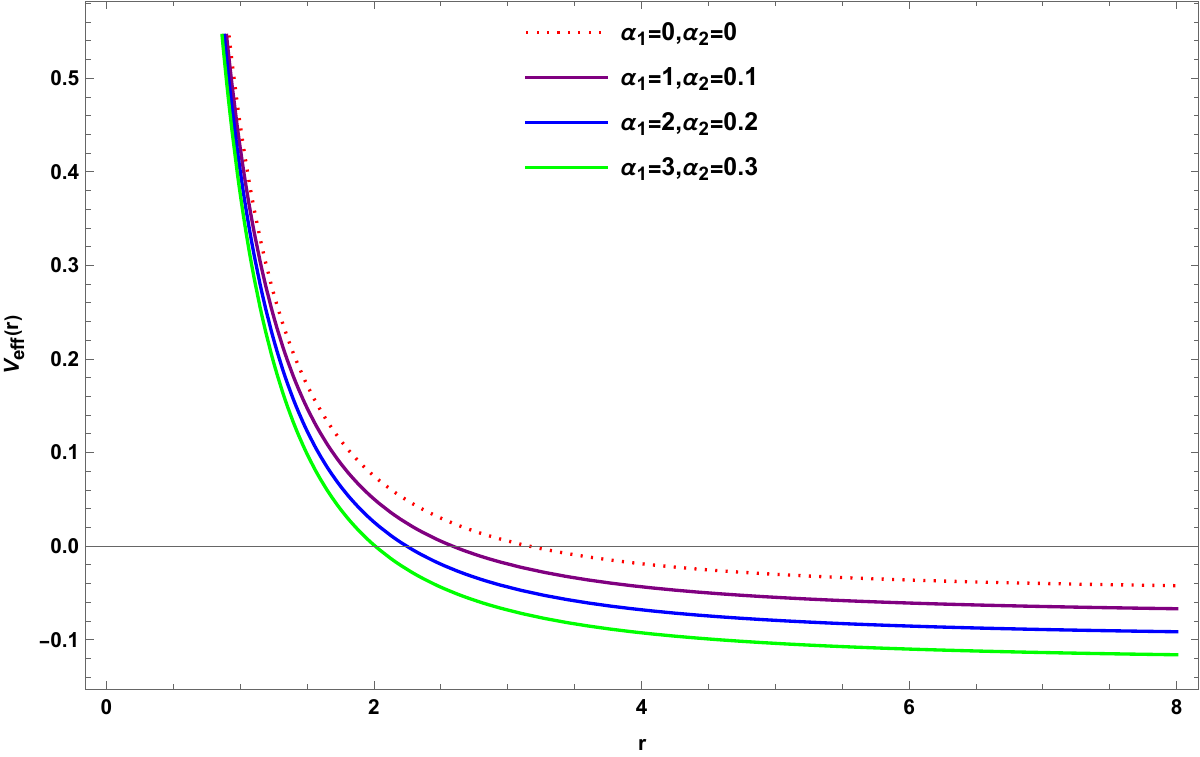}\quad\quad \includegraphics[width=0.48\linewidth]{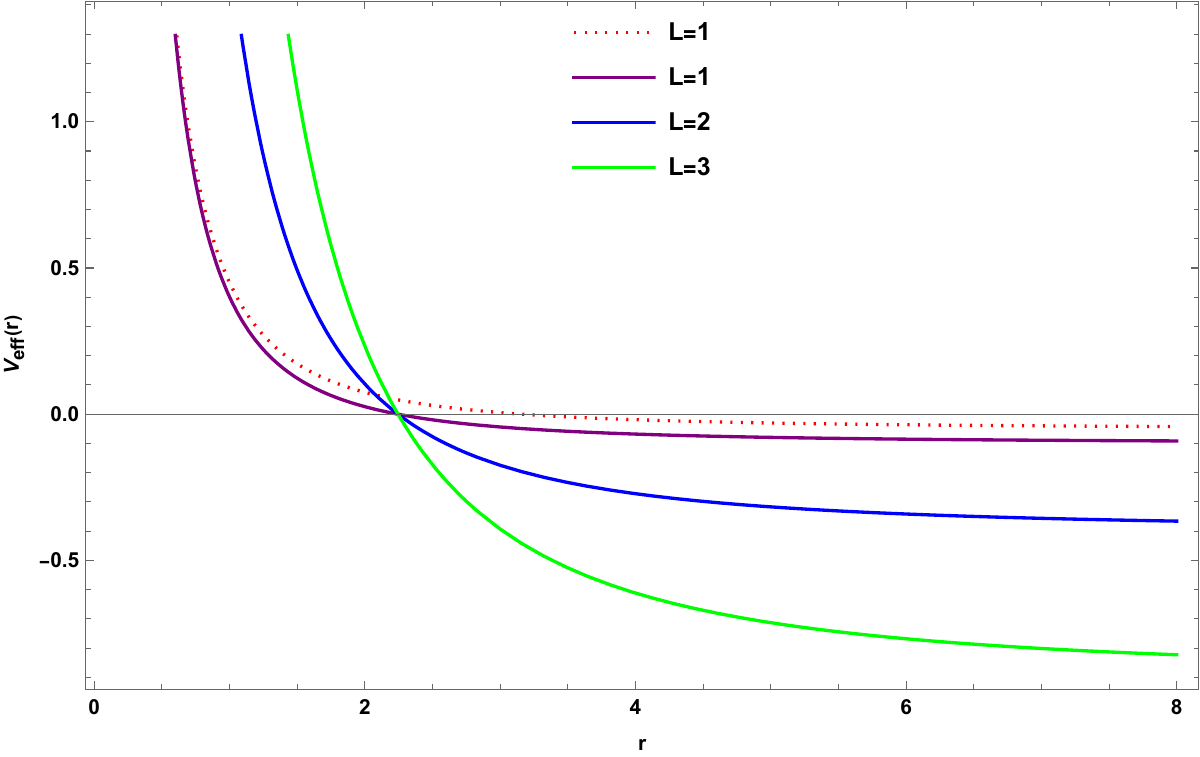}
    \caption{The potential for null geodesics in static BTZ space-time in $f(\mathcal{R})$ gravity paradigm. Here,  $\Lambda=-0.1$ and $M=1$.}
    \label{fig:8}
\end{figure}

Following are the equations for $r$ as a function of $s$ and $t$ 
\begin{equation}
\left( \frac{dr}{ds}\right) ^{2}+2V_{eff}\left( r\right) =E^{2},
\label{req1}
\end{equation}%
\begin{equation}
\left( \frac{dr}{dt}\right) ^{2}=\left( 1-\frac{2}{E^{2}}V_{eff}\left(
r\right) \right) \left( M+\Lambda _{m}r^{2}\right) ^{2}.
\end{equation}%

The effective potential $V_{eff}\left( r\right) $ of the system can be defined as 
\begin{equation}
V_{eff}\left( r\right) =\frac{1}{2}\left( M+\Lambda _{m}r^{2}\right) \left(
\epsilon +\frac{L^{2}}{r^{2}}\right) ,
\end{equation}%
where $\epsilon =0$ for null and $\epsilon =1$ for time-like geodesic.\\ It is clear that the parameters of the modified theories play an important role in describing the behavior of the effective potential. In other words, different parameter choices can result in a wide range of available behavior depending on the form of the effective potential.

\begin{figure}[htbp]
   \includegraphics[width=0.48\linewidth]{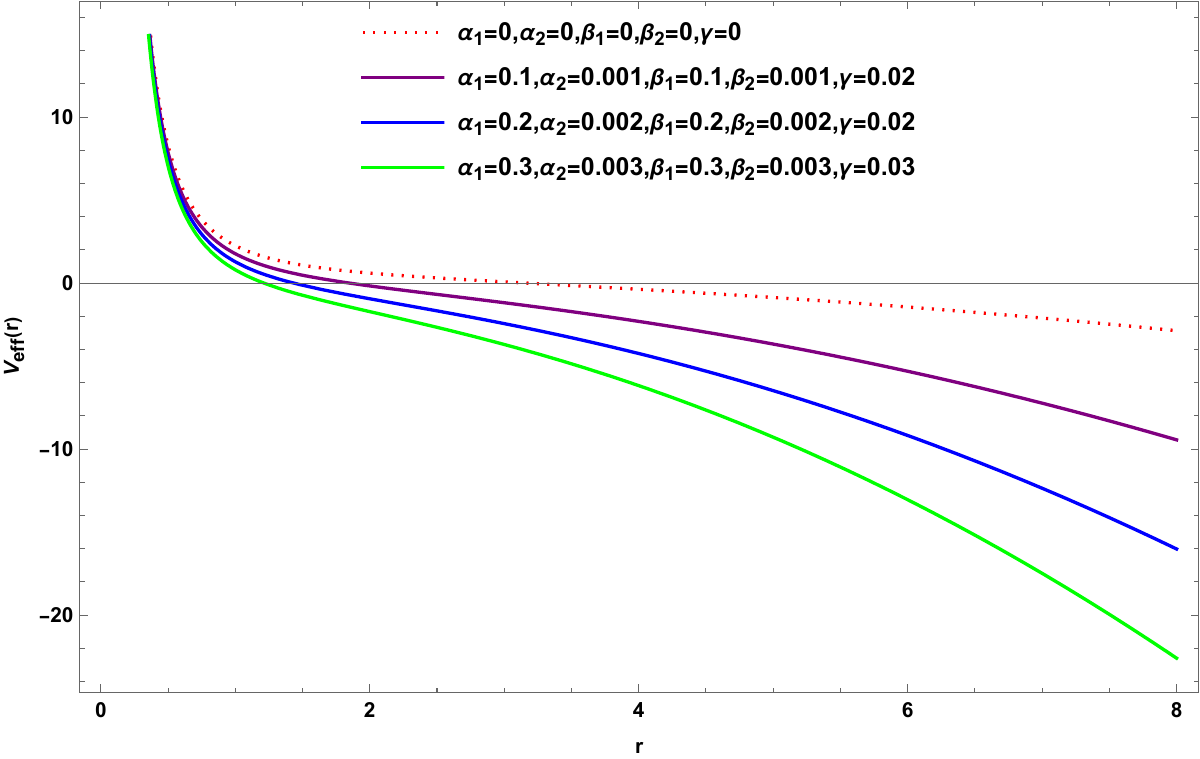}\quad\quad \includegraphics[width=0.48\linewidth]{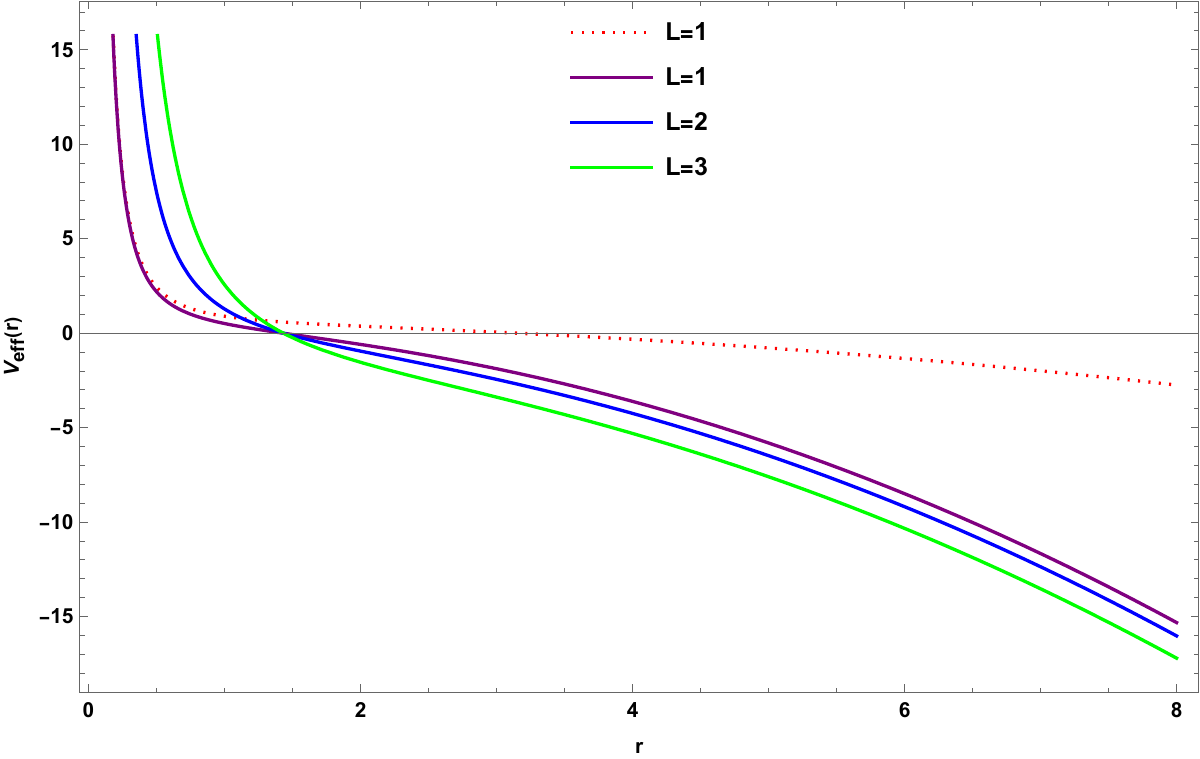}
    \caption{The potential for timelike geodesics in static BTZ space-time in $RI$-gravity paradigm. Here, $\Lambda=-0.1$ and $M=1$.}
    \label{fig:9}
\hfill\\
   \includegraphics[width=0.48\linewidth]{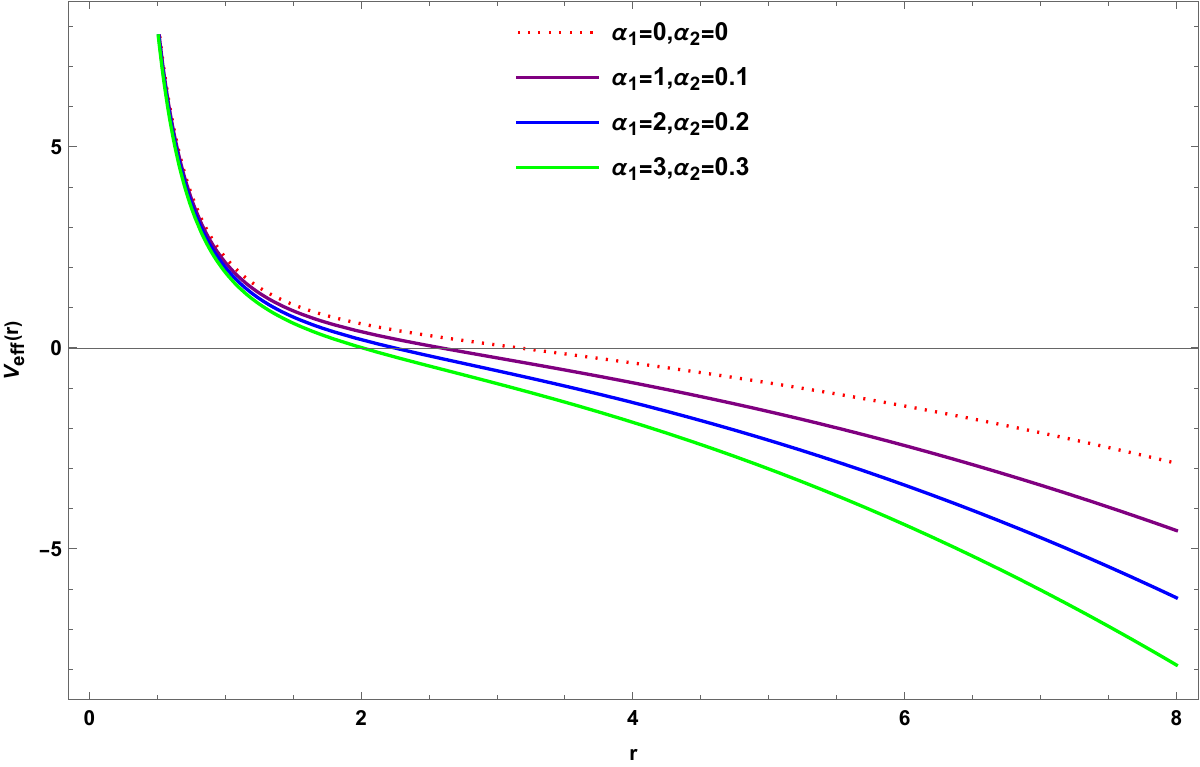}\quad\quad \includegraphics[width=0.48\linewidth]{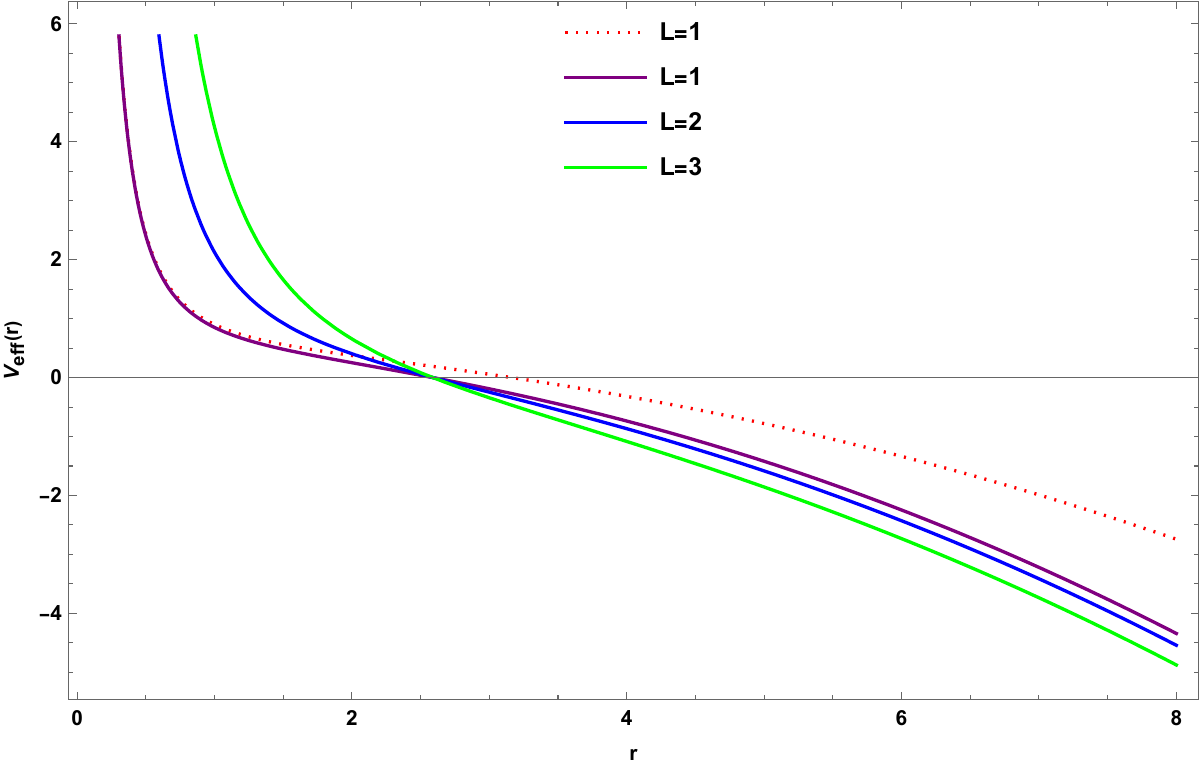}
    \caption{The potential for timelike geodesics in static BTZ space-time in $f(\mathcal{R})$-gravity paradigm. Here, $\Lambda=-0.1$ and $M=1$.}
    \label{fig:10}
\end{figure}

To investigate the effects of various factors, we examine the behavior of the potential for different parameter values. Figure \ref{fig:7} illustrates the null geodesic trajectories in both $\mathcal{RI}$-gravity and GR. In the left panel of Figure \ref{fig:7}, the dotted red line represents the GR case, where the parameters $\alpha_i = 0$, $\beta_i = 0$, $\gamma = 0$, and $L = 2$. The colored lines correspond to different values of $\alpha_i$, $\beta_i$, and $\gamma$, while maintaining the same value of $L = 2$. In the right panel of Figure \ref{fig:7}, the dotted red line again represents the GR case, and the colored lines correspond to the $\mathcal{RI}$-gravity scenario with varying values of $L$, while keeping the coupling constants fixed: $\alpha_1 = 0.2$, $\alpha_2 = 0.002$, $\beta_1 = 0.2$, $\beta_2 = 0.002$, and $\gamma = 0.02$.

In Figure \ref{fig:8}, we show the behavior of the effective potential for null geodesics in both $f(\mathcal{R})$ gravity and GR. In the left panel of Figure \ref{fig:8}, the dotted red line represents the GR case, where $\alpha_i = 0$ and $L = 1$. The colored lines correspond to different values of $\alpha_i$ in $f(\mathcal{R})$ gravity. In the right panel of Figure \ref{fig:8}, the dotted red line again represents the GR case, while the colored lines correspond to different values of $L$ in the $f(\mathcal{R})$ gravity scenario. The coupling constants are fixed as $\alpha_1 = 2$, and $\alpha_2 = 0.2$.

Figure \ref{fig:9} illustrates the behavior of time-like geodesics in both $\mathcal{RI}$-gravity and GR. In the left panel of Figure \ref{fig:9}, the red dotted line represents the GR case, where $\alpha_i = 0$, $\beta_i = 0$, $\gamma = 0$, and $L = 2$. The colored lines correspond to different values of the coupling constants $\alpha_i$, $\beta_i$, and $\gamma$ in $\mathcal{RI}$-gravity keeping the same $L=2$. In the right panel of Figure \ref{fig:9}, the red dotted line again represents the GR case, while the colored lines correspond to different values of $L$, with the coupling constants fixed as $\alpha_1 = 0.2$, $\alpha_2 = 0.002$, $\beta_1 = 0.2$, $\beta_2 = 0.002$, and $\gamma = 0.02$. 

In Figure \ref{fig:10}, we show the behavior of the effective potential for time-like geodesics in both $f(\mathcal{R})$ gravity and GR. In the left panel of Figure \ref{fig:10}, the red dotted line represents the GR case, where $\alpha_i = 0$ and $L = 2$. The colored lines correspond to different values of $\alpha_i$, while keeping $L = 2$ fixed. In the right panel of Figure \ref{fig:10}, the red dotted line, corresponding to $L = 1$, represents the GR case. The colored lines show the behavior in $f(\mathcal{R})$ gravity for different values of $L$, with the coupling constants fixed as $\alpha_1 = 2$ and $\alpha_2 = 0.2$.

Next step is to examine the geodesics of particles in the static BTZ BH
space-time. Thus, Eq.  (\ref{req1}) can be written as 
\begin{equation}
\left( r\dot{r}\right) ^{2}=r^{2}\left( E^{2}-\Lambda _{m}L^{2}- r^2 \Lambda _{m}\epsilon -M \epsilon \right)
-ML^{2}.  \label{rneq}
\end{equation}
If we consider null geodesics ($\epsilon=0$) then the analytic solution of Eq. (\ref{rneq}) is given by 
\begin{equation}
r(s)=\frac{\left( E^{2}r^{2}-L^{2}(M+\Lambda _{m} r^{2})\right) ^{3/2}}{3\left(
E^{2}-\Lambda _{m} L^{2}\right) }.
\end{equation}
On the other hand, if we consider the timelike geodesics ($\epsilon=1$) then Eq. (\ref{rneq}) has the following analytical solution 
\begin{equation*}
r(s)=\frac{M-E^2+(L^2+2\,r^2)\,\Lambda _{m}\,\sqrt{\mathrm{K}}}{8\,\Lambda
_{m}}+
\end{equation*}%
\begin{equation}
i\,\frac{\left\{E^4+(M-L^2\,\Lambda _{m})^2-2\,E^2\,(
M+L^2\,\Lambda _{m})\right\}\,\ln\left[-i\,\frac{M-E^2+(
L^2+2\,r^2)\,\Lambda _{m}}{\sqrt{\Lambda _{m}}}+2\,\sqrt{\mathrm{K}}\right]}{8\Lambda
_{m}^{3/2}},
\end{equation}%
where 
\begin{equation}
\mathrm{K}=-L^{2}\left( M+\Lambda _{m}r^{2}\right) -r^{2}\left( M+\Lambda
_{m}r^{2}-E^{2}\right) .
\end{equation}

\section{Conclusions}\label{sec:5}

In this study, we investigated the QNMs and the effective potential of massless BTZ black holes within the frameworks of $\mathcal{RI}$-gravity and $f(R)$-gravity theories, which yield the same line-element \eqref{a1} with the metric tensor (\ref{special-metric2}) and (\ref{ss4}), respectively but have different $\Lambda_{m}$. By deriving the wave equations for perturbations, we examined the behavior of the effective potential $V_{\text{eff}}(r)$ across various parameter settings, highlighting the influence of mass $m$, cosmological constant $\Lambda_{m}$, and the parameters $\alpha_1$, $\alpha_2$, $\beta_1$, $\beta_2$, and $\gamma$ in each modified gravity framework. Additionally, we analyzed the GUP-modified thermodynamics of rotating BTZ black holes in $f(R)$-gravity, offering a novel perspective on how GUP impacts the radiation of the rotating BTZ black hole in modified gravity theories. 

In Sec.~\ref{sec:3}, we examined the impact of the GUP on the thermodynamic properties of rotating BTZ black holes under the frameworks of $\mathcal{RI}$ and $f(R)$ modified gravity theories. Through the application of GUP, we derived modifications to the Hawking temperature and analyzed tunneling probabilities for massive bosons and fermions emitted from the black hole. Our findings indicate that GUP introduces a correction to the Hawking temperature, effectively reducing the black hole’s radiation rate, which has implications for the stability and evaporation of BTZ black holes in these modified gravity settings. Notably, the resulting GUP-modified Hawking temperatures vary with the parameters $\alpha_i$, $\beta_i$, and $\gamma$ in $\mathcal{RI}$-gravity, and $\alpha_i$ in $f(R)$ gravity, underscoring the significant influence of these coupling constants. These results contribute to a deeper understanding of black hole thermodynamics in quantum gravity regimes, suggesting observable deviations from GR that could be tested in high-energy astrophysical phenomena.

The visual data presented in the manuscript reveal significant patterns in the behavior of the effective potential. For example, Fig. \ref{fig:1} and Fig. \ref{fig:2} demonstrate the dependence of $V_{\text{eff}}(r)$ on the mass $m$ and parameter values in both $\mathcal{RI}$-gravity and $f(R)$-gravity, specifically for massless BTZ space-time. These figures show that the effective potential grows with an increase in $m$, indicating that the mass of the perturbing field plays a critical role in the stability of the black hole. Higher values of $m$ contribute to a more pronounced potential barrier, suggesting a greater confinement effect for perturbations. In Figs. \ref{fig:3} and \ref{fig:4}, we focused on AdS-type BTZ black holes, exploring how varying the parameters $\alpha_1$, $\alpha_2$, $\beta_1$, $\beta_2$, and $\gamma$ affects $V_{\text{eff}}(r)$. The results reveal that increasing these parameters elevates the potential barrier, thereby suggesting a stronger confinement for perturbing fields. This heightened confinement could influence the decay rates of QNMs, potentially leading to observable differences in the QNM frequencies of black holes in modified gravity frameworks. Figures \ref{fig:5} and \ref{fig:6} extend these observations by examining non-rotating BTZ black holes with varying mass and cosmological constant values. For instance, Fig. \ref{fig:5}(a) illustrates that for positive mass values ($M=2$), the effective potential forms deeper wells compared to negative mass values ($M=-2$), as seen in Figs. \ref{fig:5} (c-d) and \ref{fig:6} (c-d). This variation implies different stability characteristics for black holes depending on the sign and magnitude of the mass, highlighting the intricate interplay between mass and modified gravity parameters in shaping effective potential profiles and thereby influencing black hole stability.

Further insights are provided by examining null and timelike geodesics in static BTZ space-time, as shown in Figs. \ref{fig:7}-\ref{fig:10}. Figures \ref{fig:7} and \ref{fig:8} display the potential for null geodesics, with the potential asymptotically approaching constant values as $r$ increases. The influence of parameters $\alpha_1$, $\alpha_2$, $\beta_1$, $\beta_2$, and $\gamma$ on $V_{\text{eff}}(r)$ becomes apparent, particularly in the $RI$-gravity and $f(R)$-gravity paradigms, where higher values of these parameters significantly shift the potential profile. For timelike geodesics, as shown in Figs. \ref{fig:9} and \ref{fig:10}, we observed a sharp decline in the potential with increasing radial distance $r$, particularly with larger values of the parameters $\alpha_1$ and $\alpha_2$. This trend suggests that modified gravity parameters impact the motion of massive particles differently compared to massless ones, leading to potential observable differences in black hole environments within these theories.

Moreover, in Sec. \ref{sec:2.2}, we explored the QNMs of $AdS_3$-type BTZ black holes, analyzing how the inclusion of $\mathcal{RI}$-gravity and $f(R)$-gravity theories modifies the stability and decay rates of perturbations in this specific space-time. The modified gravity parameters induce shifts in the QNM spectra, which directly impact the damping rates of oscillations in the black hole's space-time. These findings suggest that the QNMs of $AdS_3$-type BTZ black holes could exhibit unique signatures in gravitational wave frequencies, providing a novel observational window into the effects of modified gravity theories on black hole dynamics. This analysis highlights the significance of modified gravity theories in extending our understanding of black hole perturbations, especially within an $AdS_3$ background.

In conclusion, the effective potential in BTZ space-time is shown to be significantly influenced by mass, cosmological constant, and modified gravity parameters. The variations in effective potential profiles imply differences in stability characteristics and decay rates for perturbations, thus offering deeper insights into the behavior of black holes under modified gravity frameworks. Upcoming research may expand upon these results by investigating rotating BTZ black holes and exploring alternative modified gravity models to shed more light on their influence on QNMs and the stability of black holes.

\section*{Acknowledgments}
F.A. acknowledges the Inter University Centre for Astronomy and Astrophysics (IUCAA), Pune, India for granting visiting associate-ship. \.{I}.~S. expresses gratitude to T\"{U}B\.{I}TAK, ANKOS, and SCOAP3 for their financial support. He also acknowledges COST Actions CA22113 and CA21106 for their contributions to networking.

\section*{Data Availability Statement}

This manuscript has no associated data.

\section*{Conflict of Interest}

Author(s) declare no such conflict of interests.

\section*{Code/Software}

This manuscript has no associated code/software.

\global\long\def\link#1#2{\href{http://eudml.org/#1}{#2}}
 \global\long\def\doi#1#2{\href{http://dx.doi.org/#1}{#2}}
 \global\long\def\arXiv#1#2{\href{http://arxiv.org/abs/#1}{arXiv:#1 [#2]}}
 \global\long\def\arXivOld#1{\href{http://arxiv.org/abs/#1}{arXiv:#1}}

\end{document}